\documentclass[journal]{IEEEtran}
\usepackage{booktabs}
\usepackage{cite}
\usepackage{color,soul}
\usepackage{gensymb}
\usepackage{dsfont}

\ifCLASSINFOpdf
  \usepackage[pdftex]{graphicx}
  \graphicspath{{../pdf/}{../jpeg/}}
  \DeclareGraphicsExtensions{.pdf,.jpeg,.png}
\else
  \usepackage[dvips]{graphicx}
  \graphicspath{{../eps/}}
  \DeclareGraphicsExtensions{.eps}
\fi
\usepackage[cmex10]{amsmath}

\usepackage{optidef}
\usepackage{amssymb}

\usepackage{euscript}
\usepackage{multirow}
\usepackage{algorithm}
\usepackage{algpseudocode}
\usepackage{diagbox}
\usepackage{physics}
\usepackage{tikz}
\usepackage{comment}

\ifCLASSOPTIONcompsoc
  \usepackage[caption=false,font=normalsize,labelfont=sf,textfont=sf]{subfig}
\else
  \usepackage[caption=false,font=footnotesize]{subfig}
\fi
\usepackage{url}
\usepackage{float}

\begin{document}
%
\title{A Data-Driven Game-Theoretic Approach for Behind-the-Meter PV Generation Disaggregation}
%
%
%

\author{Fankun Bu,~\IEEEmembership{Student Member,~IEEE,}
        Kaveh Dehghanpour,~\IEEEmembership{Member,~IEEE,}
        ~Yuxuan Yuan,~\IEEEmembership{Student Member,~IEEE,}
        ~Zhaoyu Wang,~\IEEEmembership{Member,~IEEE,}
        and Yingchen Zhang,~\IEEEmembership{Senior Member,~IEEE}

\thanks{This work was supported by Advanced Grid Modeling Program at the U.S. Department of Energy Office of Electricity under Grant DE-OE0000875 and the National Science Foundation under EPCN 1929975. (\textit{Corresponding author: Zhaoyu Wang})

F. Bu, K. Dehghanpour, Y. Yuan, and Z. Wang are with the Department of
Electrical and Computer Engineering, Iowa State University, Ames, IA 50011
USA (e-mail: fbu@iastate.edu; kavehdeh1@gmail.com; yuanyx@iastate.edu;
wzy@iastate.edu).

Y. Zhang is with the National Renewable Energy Laboratory, Golden, CO 80401 USA (e-mail: yingchen.zhang@nrel.gov).
}}

%
%

\markboth{}%
{Shell \MakeLowercase{\textit{et al.}}: Bare Demo of IEEEtran.cls for Journals}
%



\maketitle

\begin{abstract}
Rooftop solar photovoltaic (PV) power generator is a widely used distributed energy resource (DER) in distribution systems. Currently, the majority of PVs are installed behind-the-meter (BTM), where only customers' \textit{net demand} is recorded by smart meters. Disaggregating BTM PV generation from net demand is critical to utilities for enhancing grid-edge observability. In this paper, a data-driven approach is proposed for BTM PV generation disaggregation using solar and demand exemplars. First, a data clustering procedure is developed to construct a library of candidate load/solar exemplars. To handle the volatility of BTM resources, a novel game-theoretic learning process is proposed to adaptively generate optimal composite exemplars using the constructed library of candidate exemplars, through repeated evaluation of disaggregation residuals. Finally, the composite native demand and solar exemplars are employed to disaggregate solar generation from net demand using a semi-supervised source separator. The proposed methodology has been verified using real smart meter data and feeder models.
\end{abstract}


%
\IEEEpeerreviewmaketitle


\section{Introduction}\label{sec:intro}
\IEEEPARstart{I}{N} practice, the majority of residential rooftop PVs are installed behind-the-meter (BTM), where only the net demand is recorded, which equals native demand minus the solar power generation. Therefore, PV generation is usually invisible to distribution system operators. This invisibility, along with the stochastic nature of solar power, can cause new problems for utilities, such as inaccurate load forecasting and estimation \cite{Yi_Wang,base_load}, inefficient service restoration \cite{PV_handbook, restoration}, and sub-optimal network expansion decisions \cite{PV_detection, distribution_book, hosting_capacity}. Thus, it is of significance to disaggregate PV generation from net demand to enhance grid-edge observability. One solution to this problem is to monitor each single rooftop PV generation by installing extra metering devices. However, due to the large number of distributed PVs, this option comes at a significant cost for utilities.

To avoid costly metering infrastructure expansion, two categories of approaches have been proposed in the literature to disaggregate PV generation from net demand: \textit{Category I - Model-based methods:} Parametric models along with weather information have been used to estimate solar generation \cite{model_based_1}. In \cite{Yi_Wang}, a virtual equivalent PV station model is utilized to represent the total generation of BTM PVs in a region, where model parameters are obtained by solving an optimization problem. In \cite{SunDance}, the clear sky generation model is combined with a physical PV panel model to estimate solar generation. This model-based framework requires meteorological data, precise geographic information, and accurate physical characteristics of PV arrays. The major shortcoming of model-based solutions for solar disaggregation is the unavailability and uncertainty of model parameter information \cite{PV_cap_estimation}, which is further complicated by limited access to unauthorized BTM installations \cite{PV_detection}. Moreover, model-based solutions are subject to gross overestimation of solar generation in case of BTM PV failure \cite{PV_fault_detection}. \textit{Category II - Data-driven methods:} As the advanced metering infrastructure (AMI) has been widely deployed in distribution systems in recent years, utilities have gained access to large amounts of smart meter data \cite{smart_meter_review, the_survey}. To mine the hidden information contained within various data sources with both sufficient temporal and spatial granularity, data-driven approaches have been proposed by researchers for different applications, such as energy disaggregation \cite{disaggregation_1}, load forecasting \cite{Yi_Wang}, load management \cite{RGVP_11} and fault detection \cite{PV_fault_detection}. In particular, learning-based approaches have drawn significant attention among both researchers and industry practitioners. Measurement data from various sources, including smart meters, supervisory control and data infrastructure (SCADA), and micro-phasor measurement units ($\mu$PMU), have been utilized to perform solar disaggregation from net demand. In \cite{Hamid_Shaker}, a data-driven approach is proposed based on dimension reduction and mapping functions using PV generation measurement data from temporarily-installed sensors. In \cite{CSSS_feeder}, a linear proxy-based estimator is developed to disaggregate a solar farm generation from feeder-level measurement using $\mu$PMU data, along with the measured power profile of nearby PV plants, and global horizontal irradiance (GHI) proxy data. In \cite{indu_informatics}, a PV generation disaggregation approach is presented for groups of residential customers, under the assumption that their aggregate active power is measured at the point of common coupling (PCC) to the grid. In \cite{applied_energy}, a non-intrusive load monitoring (NILM) approach is proposed to disaggregate PV generation from net demand using measurements with 1-second resolution. In \cite{PV_cap_estimation}, the capacity of residential rooftop PVs are estimated using customers' net load curve features. In addition, several data-driven methods have been applied for disaggregating house-level demand into appliance-level energy \cite{disaggregation_1, disaggregation_2, disaggregation_3}. However, these energy disaggregation approaches also require input data with high temporal resolution in the range of a few seconds. Datasets with this level of granularity are not generally available to utilities, which applying these approaches difficult for solving solar disaggregation problem. To sum up, the chief limitations of previous data-driven methods are: dependence on installation of costly $\mu$PMU devices and high-resolution sensors throughout the network, availability of massive PV generation data, vulnerability to customer behavior volatility, and the inability to estimate all relevant parameters of BTM PV generators, including panel orientations, which impact the time-series generation profile.

Considering the drawbacks of previous methods and the emerging of smart meter data source, in this paper, a novel game-theoretic data-driven approach is proposed for disaggregating PV generation using only smart meter data. The proposed approach exploits the observed correlations within real utility data. The basic idea is to use the native demand and PV generation of fully observable customers to disaggregate the native demand and PV generation of customers with only known net demand. Accordingly, a spectral clustering (SC) algorithm is employed to construct solar and native demand \textit{candidate exemplars} using the data from fully observable customers, which are then stored in an exemplar library. Next, PV generation disaggregation is formulated as a nested bi-layer optimization problem: At the outer layer, a \textit{semi-supervised signal separation} (SSS) algorithm receives the composite native demand and solar exemplars from the inner-layer to disaggregate the native demand and PV generation from customers' net demand. The outer layer of the solar disaggregation process is subject to the response of the inner layer, at which a learning mechanism is developed to find optimal weights that are assigned to candidate native demand and solar generation exemplars to construct \textit{composite exemplars}. This mechanism is based on the concept of repeated games with vector payoff (RGVP) \cite{game_theory} in game theory. While game theory has been previously applied in power system studies \cite{RGVP_11, RGVP_22, game_theory_33}, we have not found application of RGVP theory to address the solar disaggregation challenge as presented in this paper. The learned weights are continuously updated over time using the disaggregation residual as a feedback signal. The purpose of this novel adaptive composite exemplar construction strategy is to provide optimal response to the volatile and variable behavior of customers and solar generation profiles at the grid-edge. The proposed method is validated using advanced metering infrastructure (AMI) data from our utility partners. In this paper, vectors are denoted using bold italic lowercase letters and matrices are denoted as bold uppercase letters.

The main contributions are summarized as follows:
\begin{itemize}
\item A data-driven learning-based approach is proposed for disaggregating BTM PV generation using only smart meter data. This method has been numerically compared with a model-based benchmark and has shown to have considerable improvements under incomplete information of BTM PV parameters.
\item  To find the hidden native demand and solar power values corresponding to different patterns, a closed loop game-theoretic approach is designed to learn the weights assigned to candidate exemplars for composite exemplar construction.
\item  The time-varying weights that are used for exemplar construction significantly enhance the adaptability of the disaggregator to unknown abnormal BTM events, such as PV failure and unauthorized installation and expansion of solar arrays.
\end{itemize}

The rest of the paper is organized as follows: Section \ref{sec:overall} introduces the overall framework of BTM PV generation disaggregation approach and describes smart meter dataset. Section \ref{sec:candidate_construction} proposes the method for constructing candidate exemplars. Section \ref{sec:SSS} describes the procedure of disaggregating BTM PV generation and native demand from net demand. In Section \ref{sec:game_theory}, a game-theoretic learning process is presented to obtain optimal composite PV generation and native demand exemplars. In Section \ref{sec:casestudy}, case studies are analyzed and Section \ref{sec:conclusion} concludes the paper.

\section{Proposed BTM PV Generation Disaggregation Framework and Dataset Description}\label{sec:overall}
\subsection{Overall Framework of the Proposed Approach}
In this paper, the customers are classified into three types: (I) $S_P$ denotes the set of fully observable end-users without PVs, whose native demands are directly measured by smart meters. Note that the net demand of these customers equals their native demand. (II) $S_G$ denotes the set of fully observable customers with PV generation resources. Both the native demand and the solar generation of these customers are measured separately. (III) $S_N$ represents the group of customers with BTM PVs and net demand measurements. The native demand and solar generations of these customers are unknown to the utilities. The goal of this paper is to separate aggregate BTM PV generation of groups of customers in $S_N$.

\begin{figure}[t]
\centering
\subfloat[Native demand\label{sfig:corr_demand}]{
\includegraphics[width=0.46\linewidth]{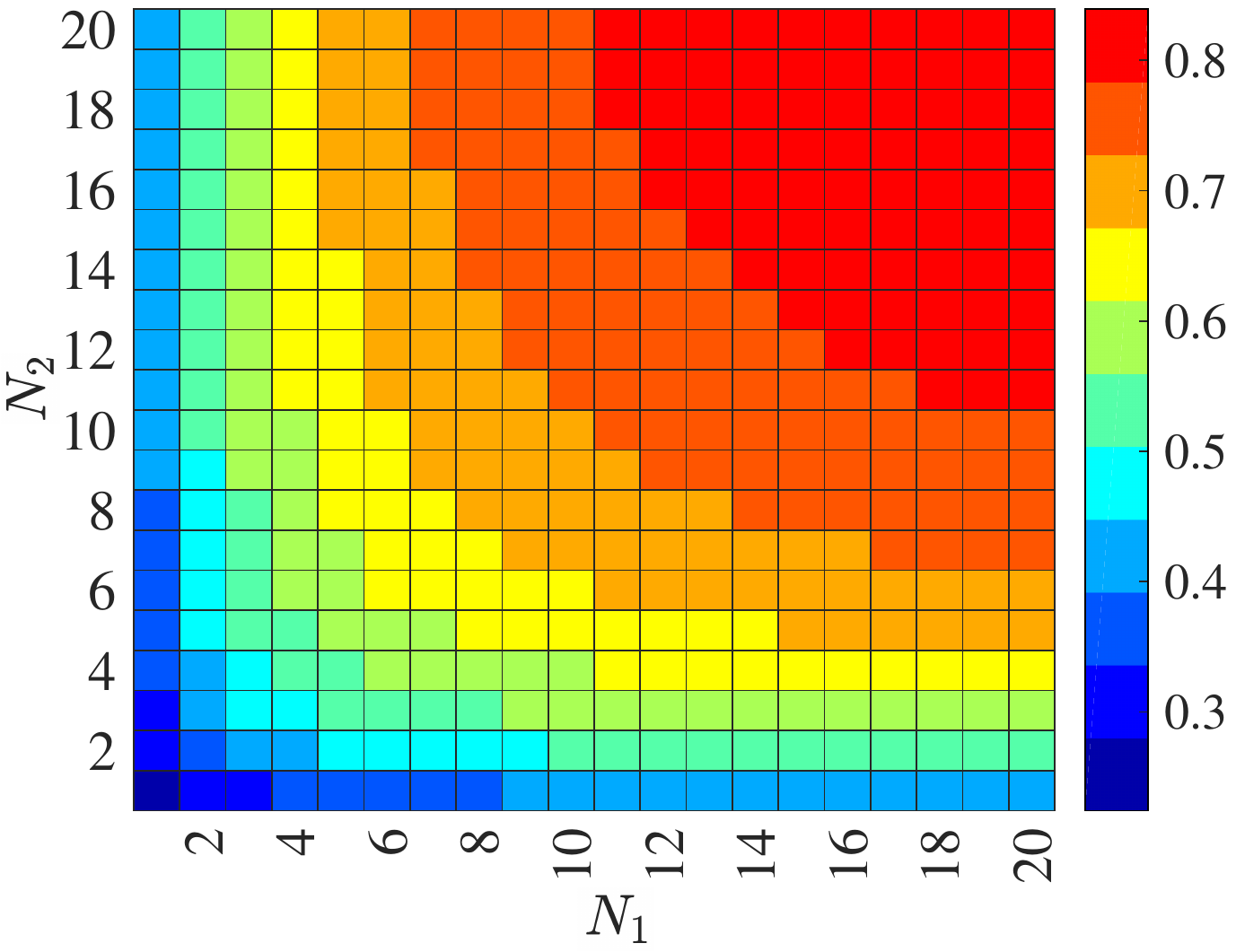}
}
\hfill
\subfloat[BTM PV generation\label{sfig:corr_PV}]{
\includegraphics[width=0.47\linewidth]{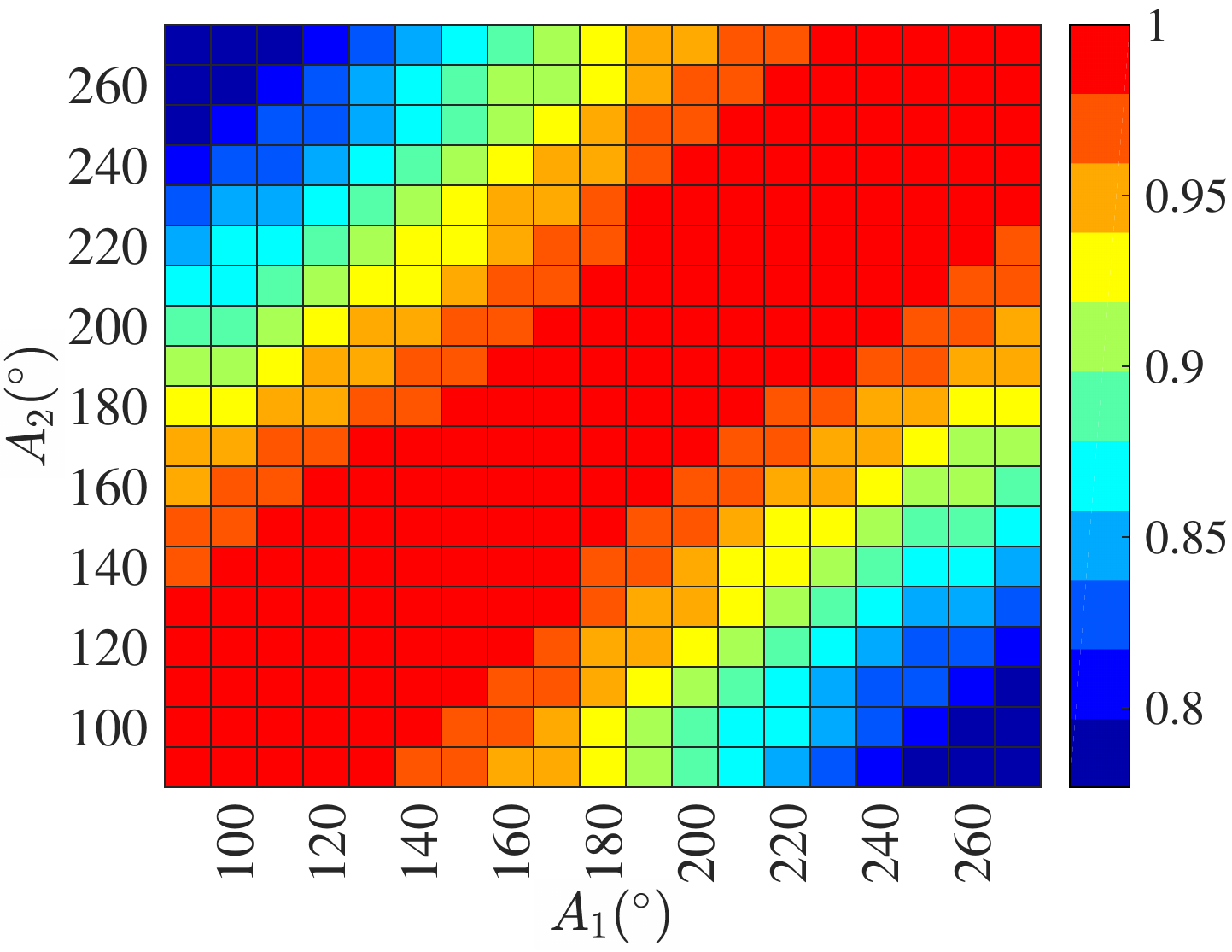}
}
\caption{Observed correlations from real smart meter data.}
\label{fig:correlation}
\end{figure}

The basic idea of the proposed BTM PV generation disaggregation approach is based on the observations that (1) the native demand of sufficiently-large groups of customers are highly correlated, (2) the PV generation of customers with similar orientation are highly correlated, and (3) the correlation between native demand and PV generation is very small. These three observations can be corroborated using real native demand and PV generation data. Fig. \ref{sfig:corr_demand} shows the correlation between the native demands of two groups of fully observable customers in $S_P$ and $S_G$, where, $N_1$ and $N_2$ denote size of each group. It can be seen that as the number of customers in each group increases, the correlation between the aggregate native demands of the customer rises as well. Fig. \ref{sfig:corr_PV} illustrates the impact of PV panel azimuth on the pairwise PV generation correlation of customers, where $A_1$ and $A_2$ denote the azimuths of two PV panels. It can be seen that as the difference between the azimuths of PV panels decreases the correlation between the solar power increases significantly. Hence, the similarity in solar generation is mainly due to similar panel orientations as expected \cite{indu_informatics, SunDance}. In contrast to the significant pairwise correlation between the native demands of groups of customers and that of BTM solar power outputs of PVs with similar orientations, the correlation between native demand and BTM PV generation is significantly small and less than $0.3$. This small correlation can be further corroborated by the mismatch between the native demand and PV generation peak times. Specifically, PV units generally output their maximum power during noon, while the native demand usually peaks in the afternoon or early evening \cite{distribution_hand_book}. The small correlation can be explained by the PVs' zero-output during nighttime, which results in a decline in the correlation between PV generation and native demand. These three observations set the foundation for constructing native demand and PV generation exemplars using the data of customers in the sets $S_P$ and $S_G$, to approximate the unobservable native demand and BTM PV generation of customers in the set $S_N$.

\begin{figure}
      \centering
      \includegraphics[width=1\columnwidth]{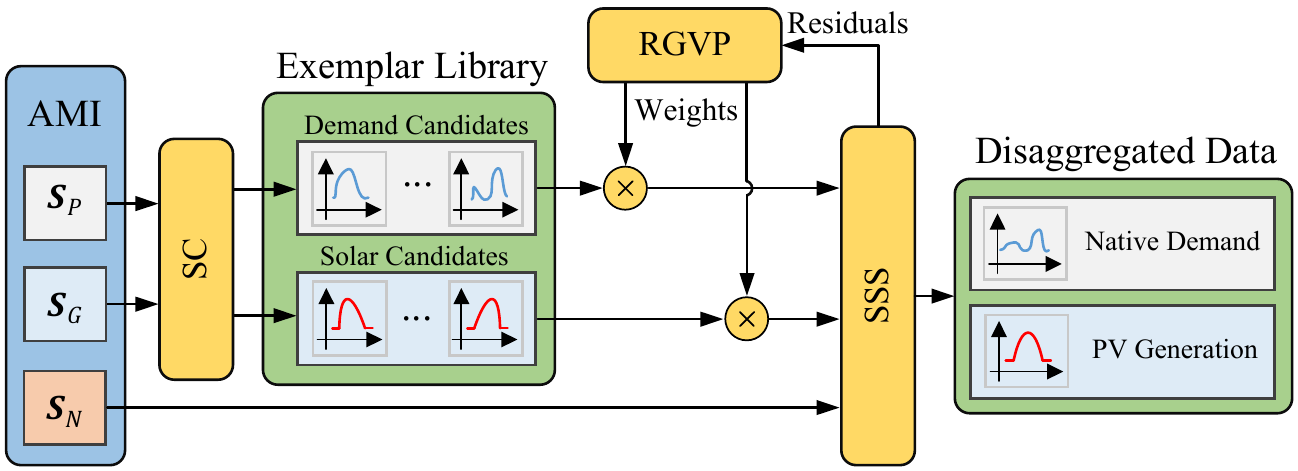}
\caption{Overall structure of the proposed BTM PV generation disaggregation method.}
\label{fig:overall}
\end{figure}

The overall disaggregation process is performed for groups of customers connected to the same lateral or secondary distribution transformer \cite{distribution_book}. The components of this process are shown in Fig. \ref{fig:overall}: \textit{Component I - Exemplar Library Construction:} The library consists of typical candidate native demand and solar generation profiles. This library is constructed based on the data of customers in the sets $S_P$ and $S_G$ using a spectral clustering (SC) algorithm. The SC algorithm automatically identifies customers with similar native load patterns and solar generation profiles. The previously-discussed correlations are used as a measure of similarity within the clustering algorithm. The outcomes of the SC method are the native demand and solar power cluster centers that are added as candidate exemplars to the library. \textit{Component II - Composite Exemplar Construction:} A weighted averaging operation is performed over the candidate native demand and PV generation exemplars within the library to generate composite native demand and solar generation exemplars. \textit{Component III - BTM PV Generation Disaggregation:} An SSS method is developed to disaggregate the net demand of customers in $S_N$ by finding the optimal coefficients that determine the share of composite native demand and solar power exemplars within the net demand. The objective of the coefficient optimization process is to minimize the disaggregation residuals. \textit{Component IV - Game-theoretic Learning:} A RGVP-based learning process is designed to assign and update the weights for each candidate exemplar in the library over time. These updated weights are then used to generate the composite native demand and solar generation exemplars for the next time point (Component II). This game-theoretic mechanism adaptively revises the behavior of the disaggregator in response to the time-varying solar power and native demand. The disaggregation process can be converted into a nested bi-layer optimization problem formulated as follows:
\begin{subequations}  \label{eq:overall}
\begin{flalign}  \label{eq:overall_a}
\underset{\pmb{p}_{t}, \pmb{g}_{t}, \alpha_{t}, \beta_{t}}{\textit{min}} \quad
 \frac{1}{2}(||\pmb{p}_t-\pmb{p}_t^C\alpha_t||_2^2+||\pmb{g}_t-\pmb{g}_t^C \beta_t||_2^2)  &&
\end{flalign}
\vspace{-20pt}
\begin{flalign}  \label{eq:overall_b}
    \quad \; \textit{s.t.} \quad \quad \; \; \pmb{p}_t+\pmb{g}_t=\pmb{p}_t^n  &&
\end{flalign}
\vspace{-20pt}
\begin{flalign}  \label{eq:overall_c}
\quad \quad\quad\quad \quad \pmb{p}_t^C = [\pmb{p}_t^{c_1},\cdots,\pmb{p}_t^{c_M}] \pmb{\omega}_t^*  &&
\end{flalign}
\vspace{-20pt}
\begin{flalign}  \label{eq:overall_d}
\quad \quad\quad\quad \quad \pmb{g}_t^C = [\pmb{g}_t^{c_1},\cdots,\pmb{g}_t^{c_N}] \pmb{\theta}_t^*  &&
\end{flalign}
\vspace{-20pt}
\begin{flalign}  \label{eq:overall_e}
\quad\quad\quad\quad\quad  \{\pmb{ \omega}_t^*, \pmb{\theta}_t^* \} & =\underset{\; \pmb{\omega}_t, \pmb{\theta}_t}{\text{argmin}} \; \Phi_{\lambda} (\pmb{p}_{t-1}^n, \mathbf{P}_t^c, \mathbf{G}_t^c, \alpha_{t-1}^*, \beta_{t-1}^*)   &&
\end{flalign}
\vspace{-18pt}
\begin{flalign}  \label{eq:overall_f}
\quad\quad\quad\quad\quad   \textit{s.t.} \quad
\sum_{i=1}^{M} \omega_{i,t} = 1, \omega_{i,t} \ge 0  &&
\end{flalign}
\vspace{-10pt}
\begin{flalign}  \label{eq:overall_g}
\quad\quad\quad\quad\quad\quad\quad \; \sum_{j=1}^{N} \theta_{j,t} = 1, \theta_{j,t} \ge 0  &&
\end{flalign}
\end{subequations}
where, $||\cdot||_2$ denotes $l_2$-norm. Note that all the demand and generation variables in this equation are defined over a time window of length $T$, where a vector $\pmb{x}_t$ represents data samples of variable $x$ in the time window $[t-T+1,t]$ as, $\pmb{x}_t=[x(t-T+1),\cdots,x(t)]$. The objective of the \textit{outer layer} is to minimize the summation of the overall disaggregation residuals, consisting of two components: 1) the difference between the actual native demand, $\pmb{p}_t$, and its alternative epitome $\pmb{p}_t^C \alpha_t$, and 2) the difference between the actual BTM PV generation, $\pmb{g}_t$, and its alternative epitome $\pmb{g}_t^C \beta_t$. Here, $\alpha_t$ and $\beta_t$ determine the proportions of demand and solar powers within the net demand, for the given composite native demand and solar exemplars, denoted by $\pmb{p}_t^C$ and $\pmb{g}_t^C$, respectively. Constraint (\ref{eq:overall_b}) ensures that the summation of the native demand and PV generation equals the observed net demand, $\pmb{p}_t^n$, which is measured by AMI. Constraints (\ref{eq:overall_c}) and (\ref{eq:overall_d}) represent the construction of composite native demand and BTM PV generation exemplars, where $\pmb{p}_t^{c_i}$ and $\pmb{g}_t^{c_j}$ are the candidate native demand and BTM PV generation exemplars, respectively. The composite exemplar construction process employs the weight vectors, $\pmb{\omega}_t=[\omega_{1,t},\cdots,\omega_{M,t}]$ and $\pmb{\theta}_t=[\theta_{1,t},\cdots,\theta_{N,t}]$, where $\omega_{i,t}$ and $\theta_{j,t}$ are the weights corresponding to candidate exemplars  $\pmb{p}_t^{c_i}$ and $\pmb{g}_t^{c_j}$, respectively. Note that each candidate exemplar represents the typical native demand/solar generation profile in the time window $[t-T+1,t]$, which are stored in the exemplar library. The objective of the \textit{inner layer}, (\ref{eq:overall_e}), is to minimize the parameterized \textit{potential function}, $\Phi_\lambda$, of the game-theoretic learning process, with parameter $\lambda$, to reduce the long-term estimation regret, where $\mathbf{P}_t^c=[\pmb{p}_t^{c_1},\cdots,\pmb{p}_t^{c_M}]$ and $\mathbf{G}_t^c=[\pmb{g}_t^{c_1},\cdots,\pmb{p}_t^{c_N}]$ are the native demand and solar generation candidate exemplar libraries, respectively. $\lambda$ is a user-defined parameter that determines the speed of updating of the weights in the game-theoretic framework (i.e., higher $\lambda$ implies faster updates). Note that the inner layer optimizes the weights at time $t$ using the measured net demand and the outcome of the outer layer at time $t-1$. Constraints (\ref{eq:overall_f}) and (\ref{eq:overall_g}) ensure that the weights assigned to the candidate exemplars are non-negative and have $l_1$-norms equal to one. The game-theoretic process assigns higher weight values to candidate exemplars that have higher impact on reducing the overall disaggregation residuals. 

\subsection{Dataset Description}
In this paper, net demand of individual customers is constructed by subtracting real BTM PV generation from real native demand. The hourly native demand and PV generation data are from Midwest U.S. utilities. The time rage of the dataset is one year and it contains 1120 customers and 337 PVs. The nominal capacity of PVs ranges from 3 kW to 8 kW. These data are available online \cite{Test_system}.

\section{Candidate Exemplar Library Construction and Composite Exemplar Generation}\label{sec:candidate_construction}
The time-series data of fully observable customers in $S_P$ and $S_G$ are leveraged to construct the candidate native demand and solar generation exemplar library using an SC method \cite{SC_tutorial, SC_book}. In performing SC, two graphs  are developed independently, each corresponding to the typical native demand and the typical PV generation libraries. The general steps in constructing the candidate exemplar libraries and composite demand/solar exemplars are as follows:

\textit{Step I - Developing similarity graphs:} The basic idea of SC technique is to solve the clustering problem using graph theory. To achieve this, a similarity graph, $\mathbf{G}=(\mathbf{V},\mathbf{E})$, is developed using AMI data.  $\mathbf{V}$ denotes the set of vertices of the graph and $\mathbf{E}$ is the set of edges connecting vertices. In our work, for the native demand, the average daily load profiles of customers in set $S_P$ are defined as the graph vertices. For PV generation, the normalized solar power profiles of the PVs in set $S_G$ are defined as graph vertices. The basic idea is to connect the vertices in $\mathbf{V}$ that are similar to each other. We have used a Gaussian kernel function as a measure of similarity, as shown below:
\begin{equation}  \label{eq:gaussian_kernel}
\mathbf{W}_{i,j}=exp(\frac{-||\mathbf{V}_i-\mathbf{V}_j||_2^2}{\rho_i \rho_j})
\end{equation}
where, $\mathbf{W}_{i,j}$ is the weight assigned to the edge connecting vertices $\mathbf{V}_i$ and $\mathbf{V}_j$, and $\rho_{i}$ and $\rho_{j}$ are tunable scaling parameters for vertices $\mathbf{V}_i$ and $\mathbf{V}_j$. The two vertices, $\mathbf{V}_i$ and $\mathbf{V}_j$, are connected when the weight of the corresponding edge, $\mathbf{W}_{i,j}$, is larger than $0$ (i.e., they have non-trivial similarities).

\textit{Step II - Developing Graph Laplacian matrices:}
Based on similarity graphs obtained from demand/solar power data, the clustering process is transformed into a graph partitioning problem, which cuts a graph into multiple smaller sections by removing edges. The graph partitioning can be conducted in different ways according to different objective functions. In this paper, the objective function is to roughly maximize the dissimilarity between the different graph clusters while minimizing the similarity within each cluster \cite{SC_tutorial}:
\begin{equation}
\label{eq:ncut1}
N_G = \min_{\mathbf{C}_1,...,\mathbf{C}_\mu}\sum_{i=1}^{\mu}\frac{\varphi(\mathbf{C}_i,\mathbf{V}\setminus{\mathbf{C}_i})}{d(\mathbf{C}_i)}
\end{equation}
where, $\mu$ is the number of clusters, $\mathbf{C}_i$ is the $i$'th cluster in graph $\mathbf{G}$, $\mathbf{V}\setminus \mathbf{C}_i$ represents the vertices of $\mathbf{V}$ that are not in $\mathbf{C}_i$, $\varphi(\mathbf{C}_i,\mathbf{V}\setminus{\mathbf{C}_i})$ represents the sum of the weights in $\mathbf{C}_i$ and $\mathbf{V}\setminus \mathbf{C}_i$, $d(\mathbf{C}_i)$ denotes the sum of the weights of the vertices in $\mathbf{C}_i$. It has been shown that the partitioning problem (equation (\ref{eq:ncut1})) can be solved using the eigenvectors of the normalized Graph Laplacian matrix \cite{SC_tutorial}, $\mathbf{L}$, as a reduced-order representation of the original data. The Laplacian is obtained as follows:
\begin{equation}  \label{eq:nor_G_Lap}
\mathbf{L}=\mathbf{D}^{-\frac{1}{2}}\mathbf{W}\mathbf{D}^{-\frac{1}{2}}
\end{equation}
where, $\mathbf{D}$ is a diagonal matrix whose diagonal elements equal the sum of elements in each row of $\mathbf{W}$. To obtain Laplacian eigenvalues, $\{\mu_1,\mu_2,...,\mu_n\}$, and the corresponding eigenvectors, eigen-decomposition is performed on $\mathbf{L} \in \mathbb{R}^{n \times n}$. The first $k$ eigenvectors corresponding to the first $k$ largest eigenvalues are concatenated into a new matrix $\mathbf{E}\in \mathbb{R}^{n \times k}$.

\textit{Step III - Obtaining candidate exemplars:}
The matrix $\mathbf{E}$ can be considered as the new representation of dataset, which embeds vertices in a lower-dimension space. It has been shown that this new matrix improves the cluster-properties of the data \cite{SC_tutorial}. Then, $k$-means algorithm is performed to cluster the rows of $\mathbf{E}$. To find the optimal number of clusters, the modified Hubert $\Gamma$ statistic index is adopted for calibration \cite{Hubert_statistic}. After that, customers in the sets $S_P$ and $S_G$ are classified into $M$ and $N$ clusters, respectively. The corresponding candidate native demand and solar generation exemplars, $\pmb{p}_t^{c_i}$ and $\pmb{g}_t^{c_j}$, are obtained using the cluster centers, which equal the average demand/solar powers for the customers belonging to each cluster.

\textit{Step IV - Constructing composite exemplars:}
The input weights from the RGVP module are used to build composite native demand and PV generation exemplars through an averaging process over the candidate exemplars:
\begin{subequations}  \label{eq:composite_demand_exemplar_1}
\begin{equation}  \label{eq:composite_demand_exemplar_a}
\pmb{p}_t^C=\displaystyle\sum_{i=1}^{M} \pmb{p}_t^{c_i}\omega_{i,t}
\end{equation}
\begin{equation}   \label{eq:composite_solar_exemplar_b}
\pmb{g}_t^C=\displaystyle\sum_{j=1}^{N} \pmb{g}_t^{c_j}\theta_{j,t}
\end{equation}
\end{subequations}
The weights, $\omega_{i,t}$ and $\theta_{j,t}$, are obtained from the RGVP-based learning process, which is elaborated in Section \ref{sec:game_theory}.

\section{BTM PV Generation and Native Demand Disaggregation}\label{sec:SSS}
Using the constructed composite native demand and solar generation exemplars for customers in the sets $S_P$ and $S_G$, the task of SSS is to estimate the coefficients of the composite exemplars, $\alpha_t$ and $\beta_t$, which are unknown \textit{a priori}. These coefficients determine the optimal disaggregation of the measured net demand for customers in $S_N$. This problem is formulated as a residual minimization problem:
\begin{subequations}  \label{eq:SSS_overall}
\begin{flalign}  \label{eq:SSS_overall_a}
\quad \quad \underset{\pmb{p}_{t}, \pmb{g}_{t}, \alpha_{t}, \beta_{t}}{\textit{min}} \quad \frac{1}{2}(||\pmb{p}_t-\pmb{p}_t^C\alpha_t||_2^2+||\pmb{g}_t-\pmb{g}_t^C \beta_t||_2^2)  &&
\end{flalign}
\vspace{-20pt}
\begin{flalign}  \label{eq:SSS_overall_b}
\quad \quad \quad \, \; \textit{s.t.} \; \quad \quad \pmb{p}_t+\pmb{g}_t=\pmb{p}_t^n  &&
\end{flalign}
\end{subequations}
where, the disaggregation residual is defined using the $l_2$-norm, which yields a convex and differentiable optimization problem that can be efficiently represented as a least-squares problem, assuming that the measurement noise follows a Gaussian distribution. Leveraging the negligible correlation between the native demand and PV generation, the optimization problem can be solved efficiently in real-time using \textit{normal equations}, which are based on introducing Lagrange multipliers to the constraint (\ref{eq:SSS_overall_b}) and employing the gradient of the objective function (\ref{eq:SSS_overall_a}). This process yields the following optimal solutions:
\begin{equation}   \label{eq:solution}
\left[
\begin{array}{c}
\alpha_t^*  \\
\beta_t^*
\end{array}
\right]
=(\mathbf{X}_t^\mathsf{T}\mathbf{X}_t)^{-1}\mathbf{X}_t^\mathsf{T}\pmb{p}_t^n
\end{equation}
where, $\mathbf{X}_t=[\pmb{p}_t^C,\pmb{g}_t^C]$. Note that although $\pmb{p}_t$ and $\pmb{g}_t$ are decision variables of interest in equation (\ref{eq:SSS_overall}), they cannot be recovered explicitly, as described in \cite{CSSS}. Instead, the optimal coefficients $\alpha_t^*$ and $\beta_t^*$ can be explicitly optimized and are used to approximate the disaggregated native demand, $\pmb{\hat{p}}_t$, and BTM PV generation, $\pmb{\hat{g}}_t$, for customers in the set $S_N$ as follows:
\begin{subequations}  \label{eq:P_G_estimated}
\begin{equation}   \label{eq:P_G_estimated_a}
\pmb{\hat{p}}_t = \pmb{p}_t^C \alpha_t^*
\end{equation}
\begin{equation}    \label{eq:P_G_estimated_b}
\pmb{\hat{g}}_t = \pmb{g}_t^C \beta_t^*
\end{equation}
\end{subequations}

\section{RGVP-based Weight Learning}\label{sec:game_theory}
In practice, the weights assigned to the candidate exemplars, $\pmb{\omega}_t$ and $\pmb{\theta}_t$, are unknown \textit{a priori} due to the unobservability of real native demand and PV generation of customers in the set $S_N$. In this section, a novel adaptive game-theoretic learning process is designed to learn these weights and then to generate and update optimal composite exemplars over time. The main idea is to handle the variations and volatility of unknown native demand and BTM solar generation by minimizing the long-term disaggregation residuals.

\begin{figure}
\centering
\includegraphics[width=0.82\linewidth]{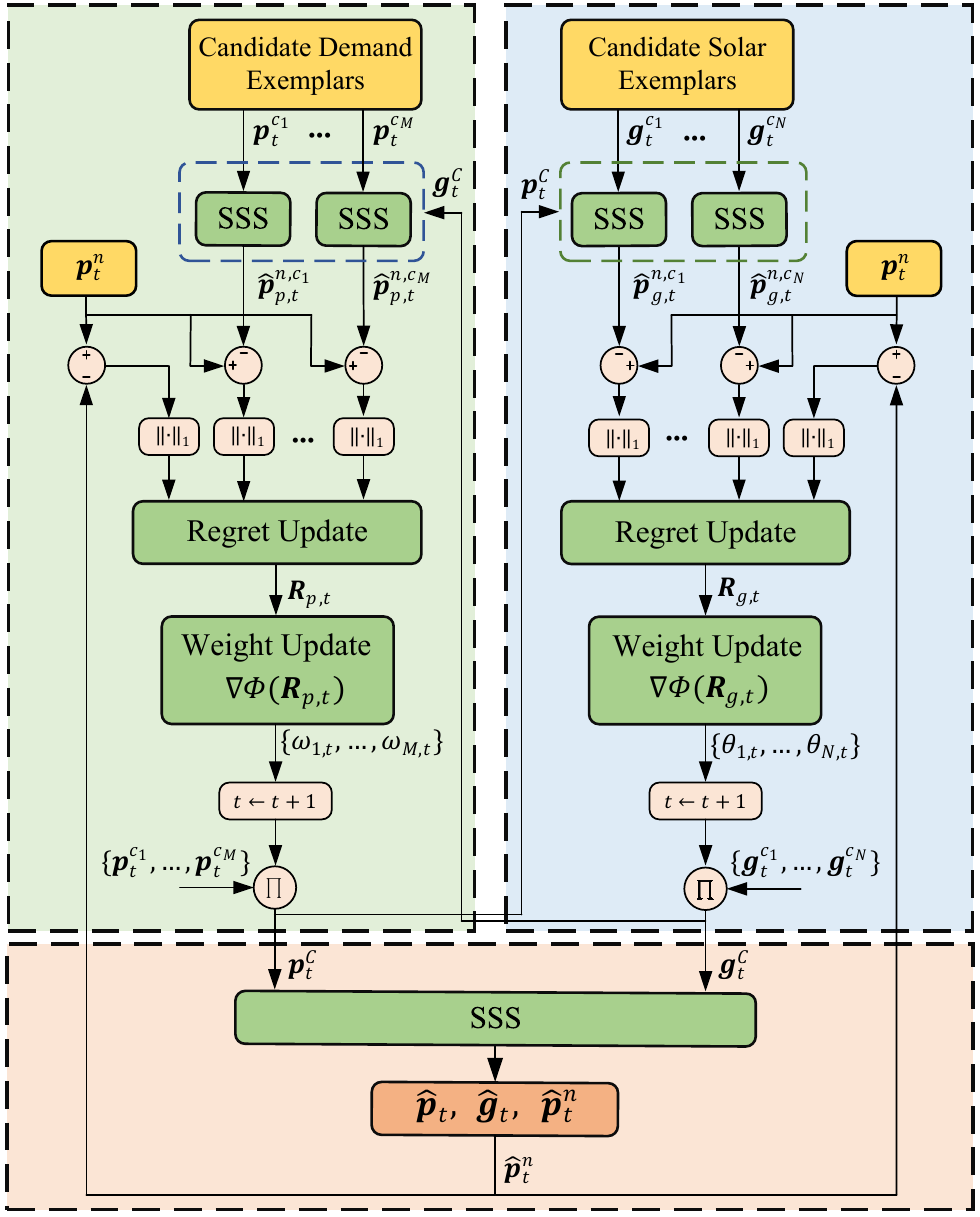}
\caption{Detailed structure of the RGVP module from Fig.2.}
\label{fig:flow_chart_game}
\end{figure}

\begin{algorithm}
\caption{BTM PV generation and native demand disaggregation from net demand}\label{alg:game_theory_virtual_code}
\begin{algorithmic}[1]
\Procedure{Initialization}{}
    \State {$t\gets t_0$, $\omega_{i,t}\gets \frac{1}{M}$, $i\in \{1,...,M\}$, $\theta_{j,t}\gets \frac{1}{N}$, $j\in \{1,...,N\}$}
\EndProcedure
\State {Receive $\{\pmb{p}_t^{c_1}, \pmb{p}_t^{c_2}, \cdots, \pmb{p}_t^{c_M}\}$ and $\{\pmb{g}_t^{c_1}, \pmb{g}_t^{c_2}, \cdots, \pmb{g}_t^{c_N}\}$ from SC}
\Procedure{Perform SSS using $\pmb{p}_t^C$ and $\pmb{g}_t^C$}{}
    \State {$\pmb{p}_t^C \gets \sum_{i=1}^{M}\pmb{p}_t^{c_i} \omega_{i,t}$, $\pmb{g}_t^C \gets \sum_{j=1}^{N}\pmb{g}_t^{c_j} \theta_{j,t}$}
    \State {$\mathbf{X}_{t} \gets [\pmb{p}_t^C,\pmb{g}_t^C]$}
    \State {$\{\alpha_t^*,\beta_t^*\} \gets (\mathbf{X}_{t}^\mathsf{T}\mathbf{X}_{t})^{-1}\mathbf{X}_{t}^\mathsf{T}\pmb{p}_t^n $}
    \State {$\pmb{\hat{p}}_t \gets \pmb{p}_t^C \alpha_t^*$, $\pmb{\hat{g}}_t \gets \pmb{g}_t^C  \beta_t^*$}
    \State {$\pmb{\hat{p}}_t^n \gets \pmb{\hat{p}}_t+\pmb{\hat{g}}_t$}
\EndProcedure
\Procedure{Perform SSS using $\pmb{p}_t^{c_i}$ and $\pmb{g}_t^C$}{}
    \State {$\mathbf{X}_{p,t}^{c_i} \gets [\pmb{p}_t^{c_i},\pmb{g}_t^C] \quad i=1,\cdots,M$}
    \State {$\{{{\alpha_{p,t}^{c_i}}^*},{{\beta_{p,t}^{c_i}}^*}\} \gets ({\mathbf{X}_{p,t}^{c_i}}^\mathsf{T}\mathbf{X}_{p,t}^{c_i})^{-1}{\mathbf{X}_{p,t}^{c_i}}^\mathsf{T}\pmb{p}_t^n $}
    \State {$\pmb{\hat{p}}_{p,t}^{c_i} \gets \pmb{p}_t^{c_i} {{\alpha_{p,t}^{c_i}}^*}$, $\pmb{\hat{g}}_{p,t}^{c_i} \gets \pmb{g}_t^C {{\beta_{p,t}^{c_i}}^*}$}
    \State {$\pmb{\hat{p}}_{p,t}^{n,c_i} \gets \pmb{\hat{p}}_{p,t}^{c_i}+\pmb{\hat{g}}_{p,t}^{c_i}$}
\EndProcedure
\Procedure{Perform SSS using $\pmb{p}_t^{C}$ and $\pmb{g}_t^{c_j}$}{}
    \State {$\mathbf{X}_{g,t}^{c_j} \gets [\pmb{p}_t^C,\pmb{g}_t^{c_j}] \quad j=1,\cdots,N$}
    \State {$\{{{\alpha_{g,t}^{c_j}}^*},{{\beta_{g,t}^{c_j}}^*}\} \gets ({\mathbf{X}_{g,t}^{c_j}}^\mathsf{T}\mathbf{X}_{g,t}^{c_j})^{-1} {\mathbf{X}_{g,t}^{c_j}}^\mathsf{T}\pmb{p}_t^n $}
    \State {$\pmb{\hat{p}}_{g,t}^{c_j} \gets \pmb{p}_t^C {{\alpha_{g,t}^{c_j}}^*}$, $\pmb{\hat{g}}_{g,t}^{c_j} \gets \pmb{g}_t^{c_j} {{\beta_{g,t}^{c_j}}^*}$}
    \State {$\pmb{\hat{p}}_{g,t}^{n,c_j} \gets \pmb{\hat{p}}_{g,t}^{c_j}+\pmb{\hat{g}}_{g,t}^{c_j}$}
\EndProcedure
\Procedure{Update Regret and Weights (Demand)}{}
    \State {$r_{p,t}^{c_i}=||\pmb{\hat{p}}_t^n-\pmb{p}_t^n||_1-||\pmb{\hat{p}}_{p,t}^{n,c_i}-\pmb{p}_t^n||_1$}
    \State {$R_{p,t}^{c_i}=\sum_{t'=t_0}^{t} r_{p,t'}^{c_i}$}
    \State {$\omega_{i,t+1} \gets e^{\lambda R_{p,t}^{c_i}}/\sum_{j=1}^{M}
    e^{\lambda R_{p,t}^{c_j}}  \quad i=1,\cdots,M$}
\EndProcedure
\Procedure{Update Regret and Weights (PV)}{}
    \State {$r_{g,t}^{c_j}=||\pmb{\hat{p}}_t^n-\pmb{p}_t^n||_1-||\pmb{\hat{p}}_{g,t}^{n,c_j}- \pmb{p}_t^n||_1$}
    \State {$R_{g,t}^{c_j}=\sum_{t'=t_0}^{t} r_{g,t'}^{c_j}$}
    \State {$\theta_{j,t+1} \gets e^{\lambda R_{g,t}^{c_j}}/\sum_{i=1}^{N}
    e^{\lambda R_{g,t}^{c_i}}  \quad j=1,\cdots,N$}
\EndProcedure
\State {$t \gets t+1$}
\State {Go to Step 4}
\end{algorithmic}
\end{algorithm}

The weight updating process is cast as a repeated game model with vector payoff, in which two components are defined: a \textit{player} and a set of \textit{experts} \cite{game_theory}, which in our problem correspond to the disaggregator and the candidate exemplars, respectively. The experts provide ``advice" (i.e., typical load/solar patterns) to the player, who then combines them to obtain the composite demand/solar exemplars. To do this, the player assigns weights to each constructed candidate exemplar and performs weighted averaging to build time-series composite exemplars, $\pmb{p}_t^C$ and $\pmb{g}_t^C$. To optimize these weights, the player determines \textit{regret} values for the advice of each expert. The regret value for each expert represents the difference between the player's loss when using the composite exemplars and the loss when using the candidate exemplar. Intuitively, a negative loss indicates how much the player is better off by performing disaggregation using composite exemplars instead of each candidate exemplar. A potential function is used to minimize the magnitude of accumulated regret vector over time by employing a gradient-based search process to optimize weight values. After that, the optimal weights are passed on to construct composite exemplars and perform the disaggregation process for customers in the set $S_N$, as described in Sections \ref{sec:candidate_construction} and \ref{sec:SSS}. The RGVP steps are shown in Fig. \ref{fig:flow_chart_game} and described as follows:

\textit{Step I - Initialization:} $t\leftarrow t_0$; uniform distribution is used to initialize the weights assigned to candidate exemplars, i.e., $\omega_{i,t}\leftarrow \frac{1}{M}$ and $\theta_{j,t}\leftarrow \frac{1}{N}$.

\textit{Step II - Construct the latest composite exemplars:} Receive candidate exemplars from exemplar libraries constructed using SC. Then, assign $\pmb{\omega}_t$ and $\pmb{\theta}_t$ to these candidate exemplars to construct composite native demand and PV generation exemplars as shown in Section \ref{sec:candidate_construction}.

\textit{Step III - Disaggregation using the latest composite exemplars:} Pass the generated composite exemplars, $\pmb{p}_t^C$ and $\pmb{g}_t^C$, to SSS (Section \ref{sec:SSS}) to perform disaggregation.  The estimated net demand is calculated using the disaggregated native demand and solar generation as follows:
\begin{equation}   \label{eq:estimated_net_demand}
\pmb{\hat{p}}_t^n=\pmb{\hat{p}}_t+\pmb{\hat{g}}_t
\end{equation}

\textit{Step IV - Determine disaggregation residual:} The disaggregation residual for the composite exemplars is obtained using the measured and estimated net demand as follows:
\begin{equation}   \label{eq:residual}
e_{t}^{C}=||\pmb{\hat{p}}_t^n-\pmb{p}_t^n||_1
\end{equation}
where, $||\cdot||_1$ denotes $l_1$-norm.

\textit{Step V - Disaggregation using the latest candidate native demand exemplars:} Instead of using composite native demand exemplar for disaggregation, candidate native demand exemplars are leveraged to perform SSS. To do this, the candidate native demand and composite PV generation exemplar pairs $\{\pmb{p}_t^{c_i}, \pmb{g}_t^{C}\}$ are passed to the SSS in parallel, $\forall i\in \{1,...,M\}$. The outcomes are the disaggregated native demand and solar generation for each pair, denoted as $\pmb{\hat{p}}_{p,t}^{c_i}$ and $\pmb{\hat{g}}_{p,t}^{c_i}$, respectively. These obtained signals are used to reconstruct the net demand corresponding to each candidate native demand exemplar, $\pmb{\hat{p}}_{p,t}^{n,c_i}$, $\forall i\in \{1,...,M\}$, as follows:
\begin{equation}   \label{eq:estimated_net_demand_candi_demand}
\pmb{\hat{p}}_{p,t}^{n,c_i} \gets \pmb{\hat{p}}_{p,t}^{c_i}+\pmb{\hat{g}}_{p,t}^{c_i}
\end{equation}
Finally, the disaggregation residual corresponding to each candidate native demand exemplar, $\pmb{p}_t^{c_i}$, is obtained as shown below:
\begin{equation}   \label{eq:residual_P}
e_{p,t}^{c_i}=||\pmb{\hat{p}}_{p,t}^{n,c_i}-\pmb{p}_t^n||_1
\end{equation}

\textit{Step VI - Disaggregation using the latest candidate solar generation exemplars:} The process introduced in Step V is performed symmetrically over candidate solar generation exemplars. Accordingly, the composite native demand and candidate PV generation exemplar pairs $\{\pmb{p}_t^{C}, \pmb{g}_t^{c_j}\}$ are passed to the SSS, $\forall j\in \{1,...,N\}$, where the disaggregated native demand and solar generation are obtained, denoted as $\pmb{\hat{p}}_{g,t}^{c_j}$ and $\pmb{\hat{g}}_{g,t}^{c_j}$, respectively. These disaggregated signals are then used to reconstruct the net demand, $\pmb{\hat{p}}_{g,t}^{n,c_j}$, corresponding to each candidate solar exemplar and determine the disaggregation residuals $e_{g,t}^{c_j}$, similar to equation (\ref{eq:residual_P}).

\textit{Step VII - Update candidate regrets:} The player's instantaneous regrets for each native demand and solar generation candidate exemplars are calculated as follows:
\begin{subequations}  \label{eq:inst_regret}
\begin{equation}
r_{p,t}^{c_i}=e_t^C - e_{p,t}^{c_i}  \quad i=1,\cdots,M
\end{equation}
\begin{equation}
r_{g,t}^{c_j}=e_t^C - e_{g,t}^{c_j}  \quad j=1,\cdots,N
\end{equation}
\end{subequations}
here, $r_{p,t}^{c_i}$ and $r_{g,t}^{c_j}$ represent the regrets for candidate demand and solar exemplars, respectively, which are measured in terms of the payoffs in disaggregation residuals by following the advice of candidate exemplars instead of composite exemplars at time $t$. The cumulative regrets for the $i$'th candidate native demand exemplar and the $j$'th candidate PV generation exemplar are defined by summing up the instantaneous regret $r_{p,t}^{c_i}$ and $r_{g,t}^{c_j}$ for all previous time instants in $[t_0,t]$, as follows:
\begin{subequations}  \label{eq:cumu_regret}
\begin{equation}
R_{p,t}^{c_i}=\displaystyle\sum_{t'=t_0}^{t} r_{p,t'}^{c_i} \quad i=1,\cdots,M
\end{equation}
\begin{equation}
R_{g,t}^{c_j}=\displaystyle\sum_{t'=t_0}^{t} r_{g,t'}^{c_j} \quad j=1,\cdots,N
\end{equation}
\end{subequations}
By assigning accumulated regret values to each expert, the regret vectors are obtained as $\pmb{R}_{p,t}=[R_{p,t}^{c_1},\cdots,R_{p,t}^{c_M}]^\mathsf{T}$ and $\pmb{R}_{g,t}=[R_{g,t}^{c_1},\cdots,R_{g,t}^{c_N}]^\mathsf{T}$ for candidate native demand and solar generation exemplars, respectively.

\textit{Step VIII - Update weights:} The goal of RGVP is to reduce the magnitude of the accumulated regret vectors $\pmb{R}_{p,t}$ and $\pmb{R}_{g,t}$. To do this, \textit{potential functions} are assigned to these accumulated vector spaces. These scalar potential functions are increasing with respect to the advisors' accumulated regrets. Hence, reducing the accumulative regret is transformed into minimizing the values of these potential functions \cite{game_theory}. In this paper, we have adopted exponential potential functions, the gradients of which are used to update the weights as follows \cite{game_theory}:
\begin{subequations}  \label{eq:weight_solution}
\begin{equation}
\omega_{i,t+1}=\nabla \Phi_\lambda (\pmb{R}_{p,t})_i=\frac{e^{\lambda R_{p,t}^{c_i}}}{\sum_{j=1}^{M}
e^{\lambda R_{p,t}^{c_j}}} \quad i=1,\cdots,M
\end{equation}
\begin{equation}
\theta_{j,t+1}=\nabla \Phi_\lambda (\pmb{R}_{g,t})_j=\frac{e^{\lambda R_{g,t}^{c_j}}}{\sum_{i=1}^{N}
e^{\lambda R_{g,t}^{c_i}}} \quad j=1,\cdots,N
\end{equation}
\end{subequations}
where,
\begin{equation}   \label{eq:exponential_potential}
\Phi_\lambda (\pmb{u})=\frac{1}{\lambda} \mathrm{ln} \Big(\displaystyle\sum_{i=1}^{L}e^{\lambda u_i}  \Big)
\end{equation}
is an exponential potential operator with $\pmb{u}=[u_1,\cdots,u_L]^\mathsf{T}$, $\lambda$ is a positive tunable parameter indicating the updating speed of weights, which is adopted from literature as $\lambda = \sqrt{8\mathrm{ln}(L)/T}$ \cite{game_theory}, with $L=M$ or $L = N$.

\textit{Step IX - Moving the disaggregation window:} $t\leftarrow t+1$; go back to Step II.

An algorithmic overview of the aforementioned steps of BTM PV generation disaggregation is summarized in Algorithm \ref{alg:game_theory_virtual_code}.

\begin{figure*}
      \centering
      \includegraphics[width=2\columnwidth]{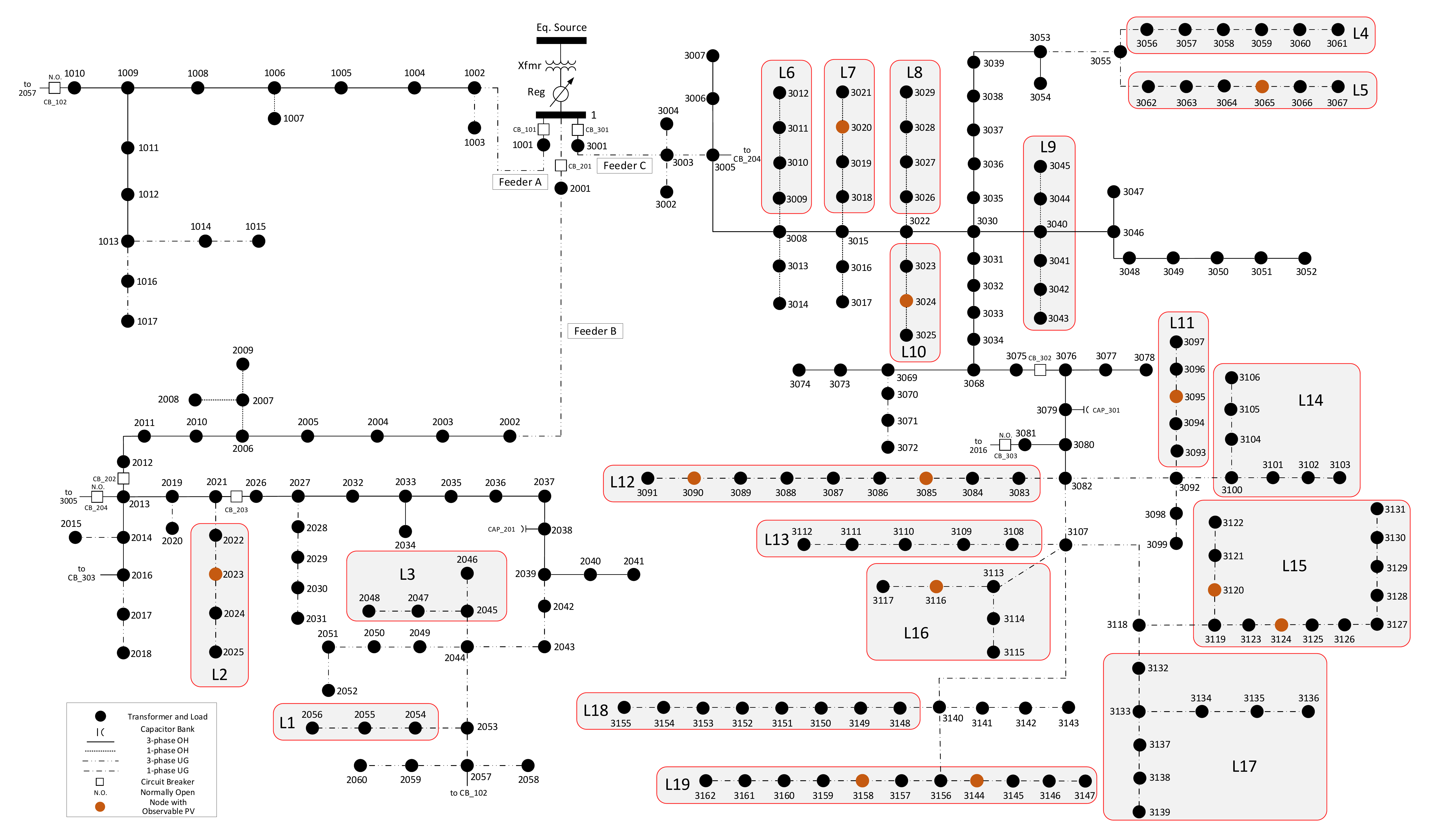}
\caption{One-line diagram of a real distribution system.}
\label{fig:overall_topology}
\end{figure*}

\section{Case Study}\label{sec:casestudy}
In this section, the proposed BTM PV generation disaggregation method is verified using real smart meter data described in Section \ref{sec:overall} and the one-line diagram of a 240-node distribution grid presented in \cite{Test_system}. The proposed approach is applied to disaggregate single-phase lateral- and transformer-level PV generation over a one-year data period. The number of customers, $N_P$, and BTM PV generators, $N_G$, connected to the system laterals are shown in Table \ref{tbl:lateral_information}.

\begin{table}[h]
	\renewcommand{\arraystretch}{1.5}
	\setlength{\tabcolsep}{4.8pt}
	\caption{Number of customers and PVs in Laterals}\label{tbl:lateral_information}
	\begin{tabular}{ccccccccccc}
	    \toprule[1pt]
		\textbf{} & \textbf{L1} & \textbf{L2} & \textbf{L3} & \textbf{L4} & \textbf{L5} & \textbf{L6} & \textbf{L7} & \textbf{L8} & \textbf{L9} & \textbf{L10} \\
	    \hline
        $N_P$ & 22 & 26 & 30 & 36 & 25 & 27 & 30 & 24 & 37 & 23\\
        $N_G$ & 13 & 13 & 21 & 21 & 18 & 16 & 14 & 9  & 20 & 14\\
        \bottomrule[1pt]
		\textbf{} & \textbf{L11} & \textbf{L12} & \textbf{L13} & \textbf{L14} & \textbf{L15} & \textbf{L16} & \textbf{L17} & \textbf{L18} & \textbf{L19} \\
	    \hline
        $N_P$ & 16 & 26 & 29 & 37 & 67 & 13 & 43 & 24 & 33\\
        $N_G$ & 10 & 14 & 24 & 17 & 42 & 9  & 23 & 14 & 25\\
    	\bottomrule[1pt]
	\end{tabular}
\end{table}

The tunable parameters in the proposed BTM PV generation disaggregation approach include the number of candidate native demand exemplars, $M$, the number of candidate PV generation exemplars, $N$, and the disaggregation time-window length, $T$. To optimize the number of clusters, $M$ and $N$, the modified Hubert $\Gamma$ statistic index is calculated for different number of clusters by running SC \cite{Hubert_statistic}. Then, the optimal values of $M$ and $N$ are determined by finding the knee point of $\Gamma$ curve as presented in \cite{determine_knee}. In our case, $M$ and $N$ are optimized at 4 and 3, respectively. To tune the length of the moving time window, a grid search method was employed to find the minimum net demand estimation residual in terms of mean absolute percentage error ($MAPE$), calculated as follows:
\begin{equation}   \label{eq:MASE_eq}
MAPE=\frac{100\%}{K}\cdot \sum_{t=1}^{K} \Bigg| \frac{\hat{p}^n(t)-p^n(t)}{\frac{1}{K} \sum_{t=1}^{K}|p^n(t)|}  \Bigg|
\end{equation}
where, $K$ is the total number of net demand samples. In our case, the identified optimal value of $T$ is 96 hours.

\begin{figure}
\centering
\subfloat[PV generation\label{sfig:disaggregated_PV}]{
\includegraphics[width=0.8\linewidth]{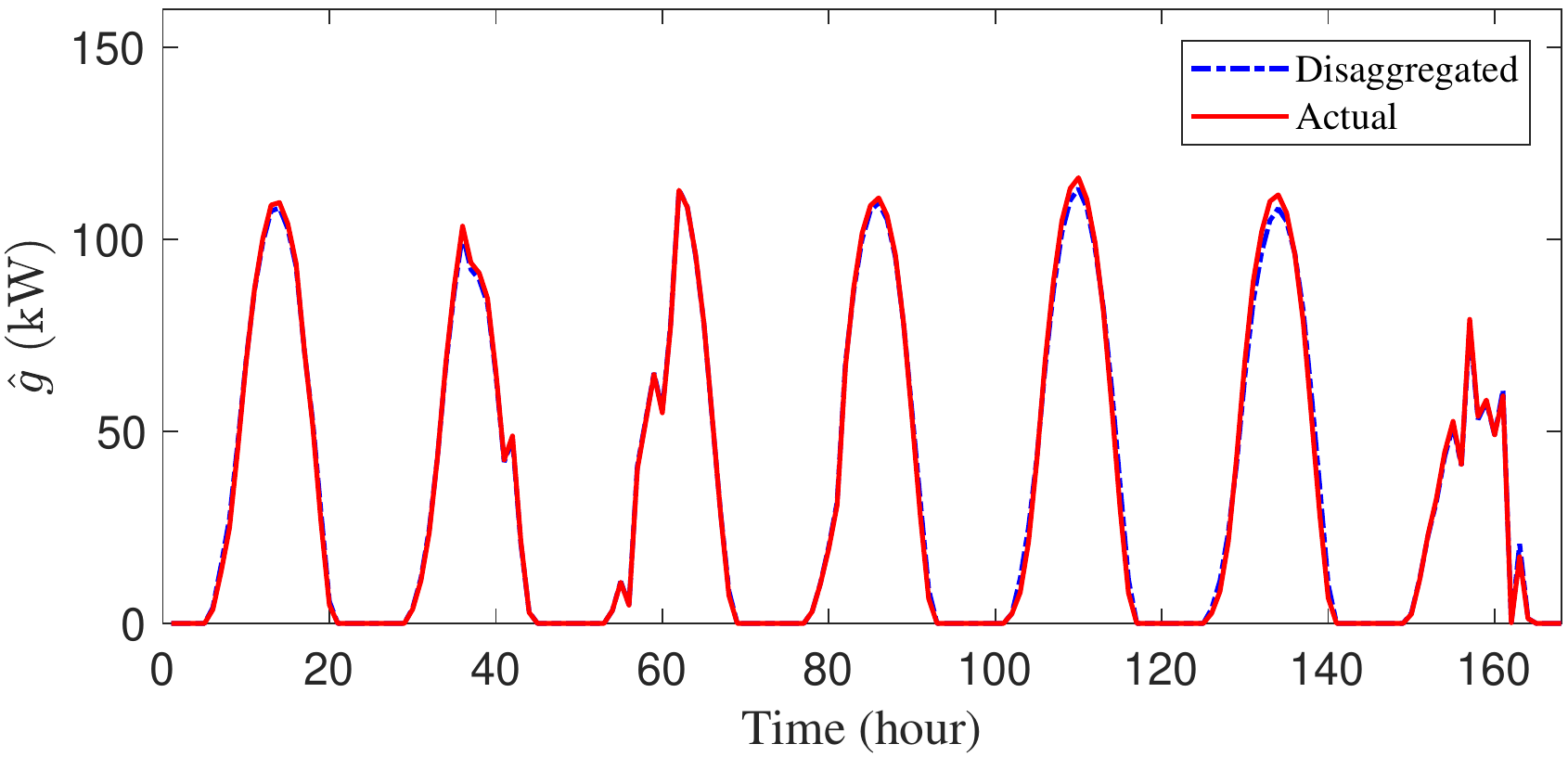}
}
\hfill
\subfloat[Native demand\label{sfig:disaggregated_demand}]{
\includegraphics[width=0.8\linewidth]{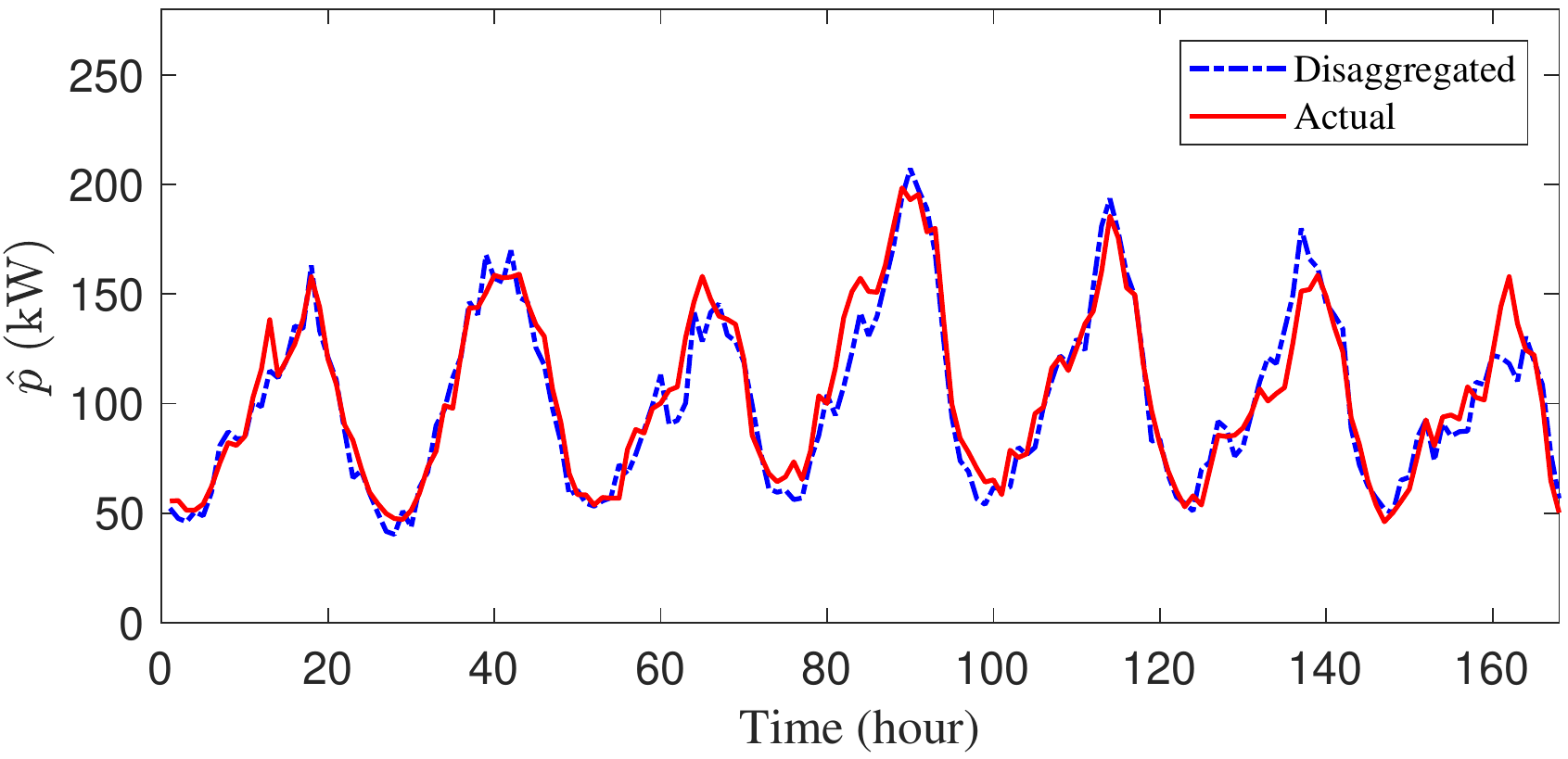}
}
\caption{PV generation and native demand profiles of a lateral within a one-week period.}
\label{fig:P_G_disaggregation}
\end{figure}

These calibrated parameters are then used to perform BTM PV generation disaggregation. Fig. \ref{fig:P_G_disaggregation} shows the disaggregated PV generation and native demand for a lateral within a one-week period. It can be seen in Fig. \ref{sfig:disaggregated_PV} that the disaggregated PV generation closely fits the actual unobservable solar power, which indicates a high disaggregation accuracy. Also, as demonstrated by the disaggregated and actual PV generation curves during the last day, despite high variations in PV generation profile, the proposed approach can still provide satisfying accuracy. The disaggregated native demand and the actual native demand are shown in Fig. \ref{sfig:disaggregated_demand}, where the disaggregated native demand can accurately capture the load variations as well.

The proposed approach has been applied to 19 laterals in a real distribution system to test its generalizability. The lateral divisions are illustrated in Fig. \ref{fig:overall_topology}, where the nodes that have observable PVs are shown with brown color. The smart meter data from individual customers are aggregated to obtain lateral-level demand profiles, in which the utilities have shown significant interest. In our case study 5\% of all customers are fully observable with known PV generation and native demand data (the set $S_G$). The remaining 95\% belong to the sets $S_N$ and $S_P$, where only either the net demand or the native demand is observable, respectively. The idea is to use the data of $S_G$ and $S_P$ to disaggregate PV generation of $S_N$. Using the actual PV generation and native demand as ground truths, the fitness of the disaggregated PV generation and native demand are evaluated in terms of $MAPE$, as shown in Table \ref{tbl:MASE_table}, where the solar and demand disaggregation accuracy are denoted as $g_{m}$ and $p_{m}$, respectively. As can be seen, $g_{m}$ ranges from 4\% to 8\%, and $p_{m}$ ranges from 6\% to 10\%. Note that in the case of solar disaggregation using feeder-level demand profile measurements, e.g., SCADA power flow data, the network losses should be removed before disaggregation. To do this, previous works have proposed several power flow-based techniques to estimate and eliminate the network losses \cite{loss_estimation, loss_estimation_1}. After this, PV generation disaggregation can be performed using our proposed method.

\begin{table}[h]
	\renewcommand{\arraystretch}{1.5}
	\setlength{\tabcolsep}{3.8pt}
	\caption{PV generation and native demand disaggregation MAPE}\label{tbl:MASE_table}
	\begin{tabular}{ccccccccccc}
	    \toprule[1pt]  
		\textbf{} & \textbf{L1} & \textbf{L2} & \textbf{L3} & \textbf{L4} & \textbf{L5} & \textbf{L6} & \textbf{L7} & \textbf{L8} & \textbf{L9}  & \textbf{L10} \\
	    \hline
        $g_m$ (\%) & 6.10 & 8.09 & 4.43 & 6.68 & 7.38 & 5.34 & 6.81 & 8.08 & 4.86  & 4.96 \\
        $p_m$ (\%) & 8.44 & 8.47 & 7.52 & 7.84 & 8.03 & 7.99 & 7.72 & 8.45 & 7.52  & 8.96 \\
	    \toprule[1pt]
		\textbf{} &  \textbf{L11} & \textbf{L12} & \textbf{L13} & \textbf{L14} & \textbf{L15} & \textbf{L16} & \textbf{L17}  & \textbf{L18}  & \textbf{L19}\\
	    \hline
        $g_m$ (\%)  & 6.33 & 6.12 & 5.79 & 6.97 & 3.66 & 5.58 & 4.32 & 4.58 & 3.44 \\
        $p_m$ (\%)  & 9.47 & 7.23 & 9.04 & 7.52 & 6.59 & 9.66 & 6.62 & 8.38 & 7.51 \\
        \bottomrule[1pt]
	\end{tabular}
\end{table}

\begin{figure}
\centering
\subfloat[Candidate solar generation weights\label{sfig:omega_PV}]{
\includegraphics[width=0.8\linewidth]{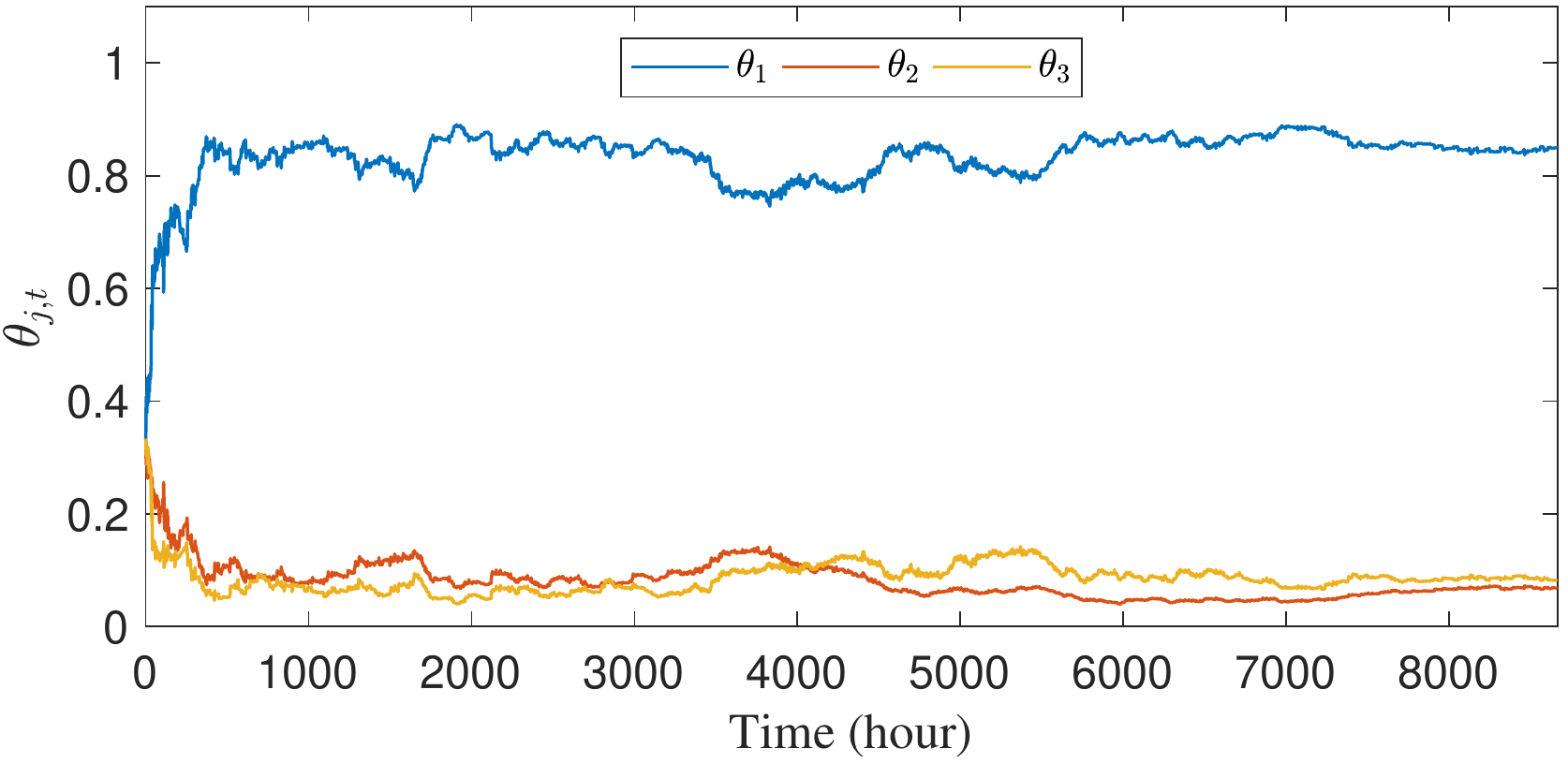}
}
\hfill
\subfloat[Candidate native demand weights\label{sfig:omega_demand}]{
\includegraphics[width=0.8\linewidth]{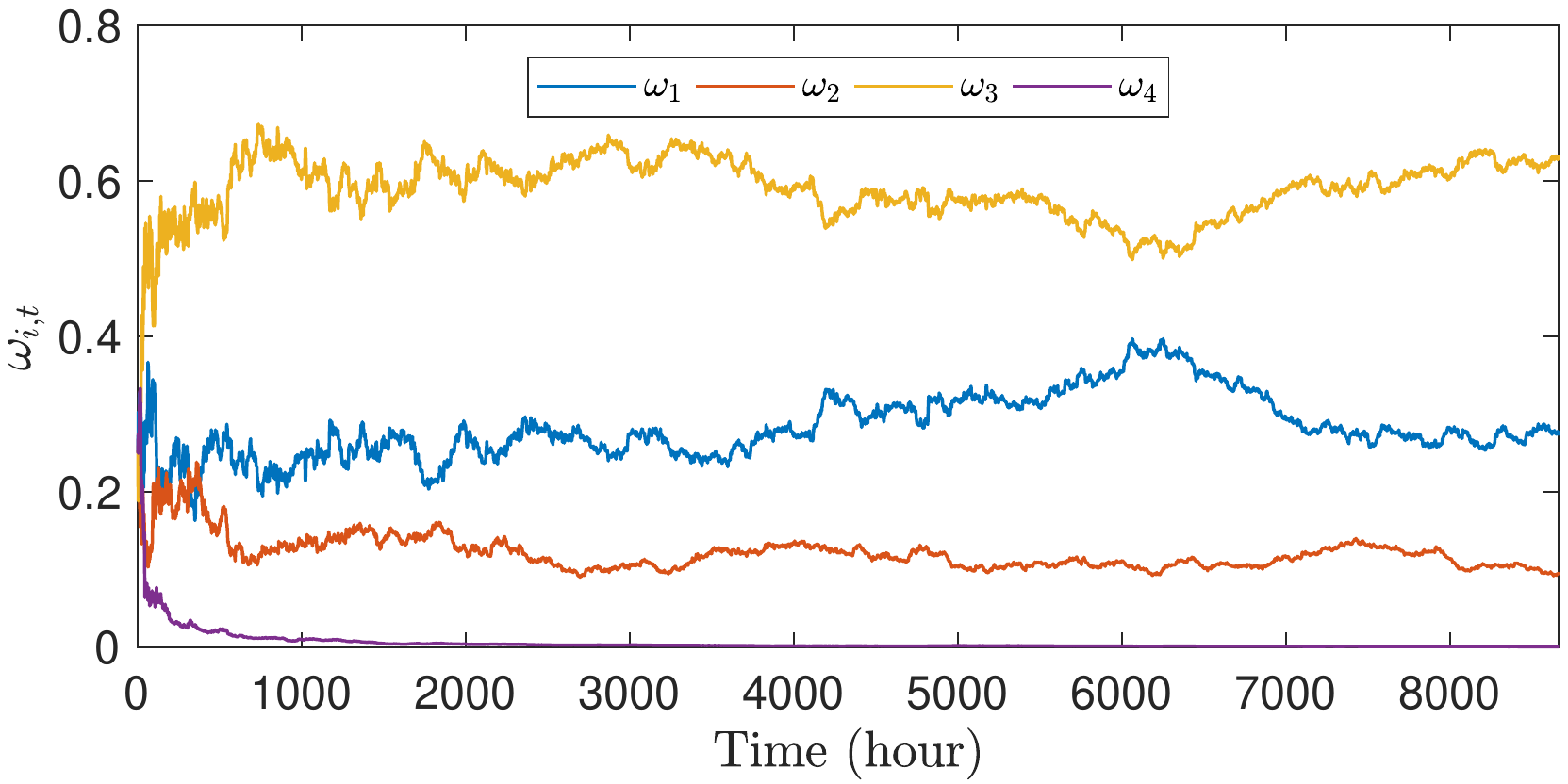}
}
\caption{The time-series weights assigned to the candidate PV generation and native demand exemplars.}
\label{fig:omega_variation}
\end{figure}

The case study is conducted on a standard PC with an Intel(R) Xeon(R) CPU running at 3.70 GHz and with 32.0 GB of RAM. PV generation disaggregation is performed for each lateral illustrated in Fig. \ref{fig:overall_topology} over a year, and the computational time ranges from 4.20 seconds to 4.64 seconds.

Furthermore, it is of interest to examine the variations of the obtained game-theoretic weights corresponding to different candidate exemplars. In Fig. \ref{sfig:omega_PV}, it can be seen that the weights assigned to the three candidate PV generation exemplars for one of the laterals converge to approximately 0.8, 0.1, and 0.1, through the learning process, after which the weight values remain nearly stable. Similar characteristics can be observed in Fig. \ref{sfig:omega_demand}, which shows the variations of weights assigned to the candidate native demand exemplars for the same lateral.

\begin{figure}[h]
\centering
\subfloat[Solar power estimation error\label{sfig:PV_generation_residual}]{
\includegraphics[width=0.8\linewidth]{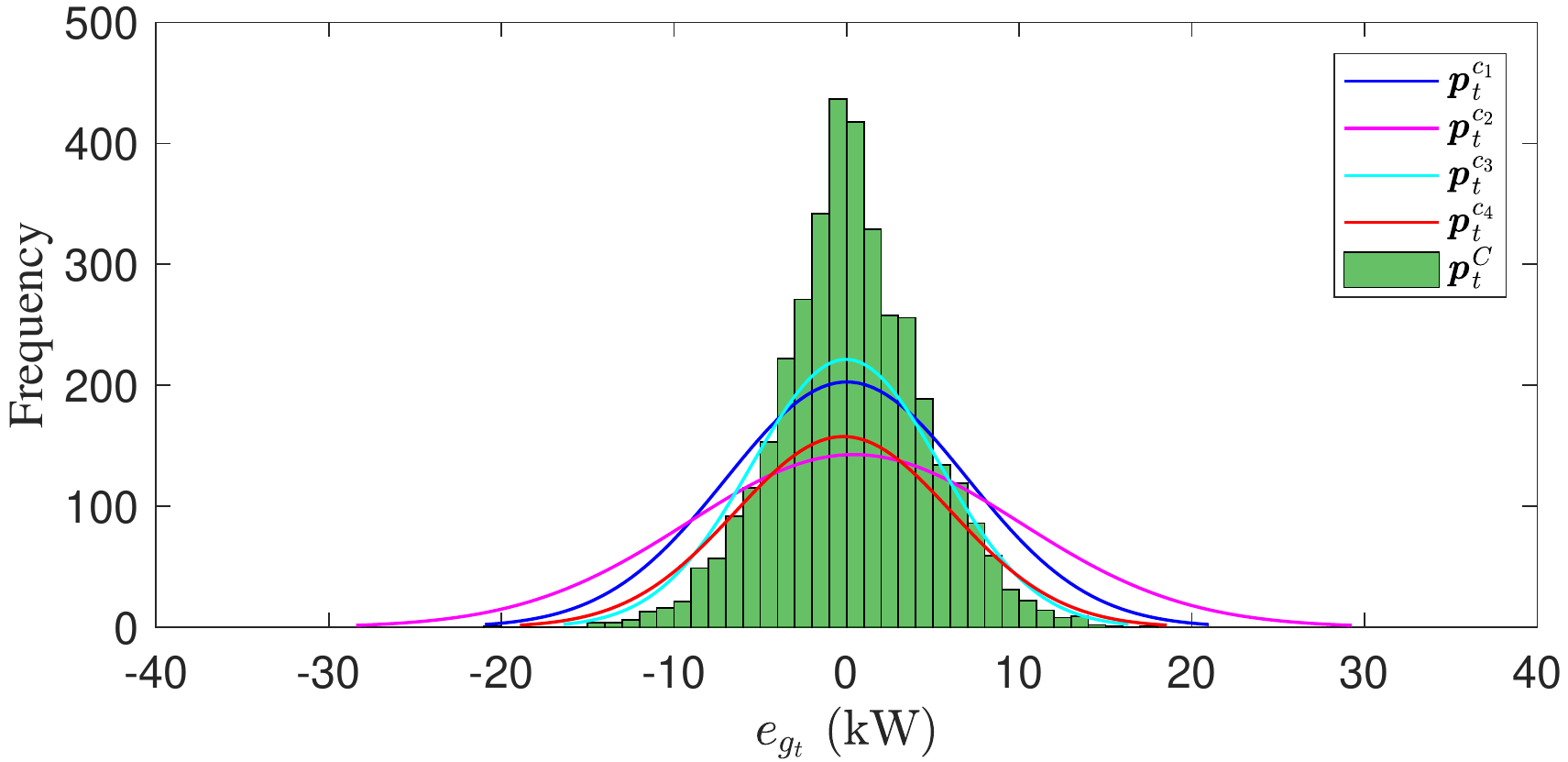}
}
\hfill
\subfloat[Native demand estimation error\label{sfig:native_demand_residual}]{
\includegraphics[width=0.8\linewidth]{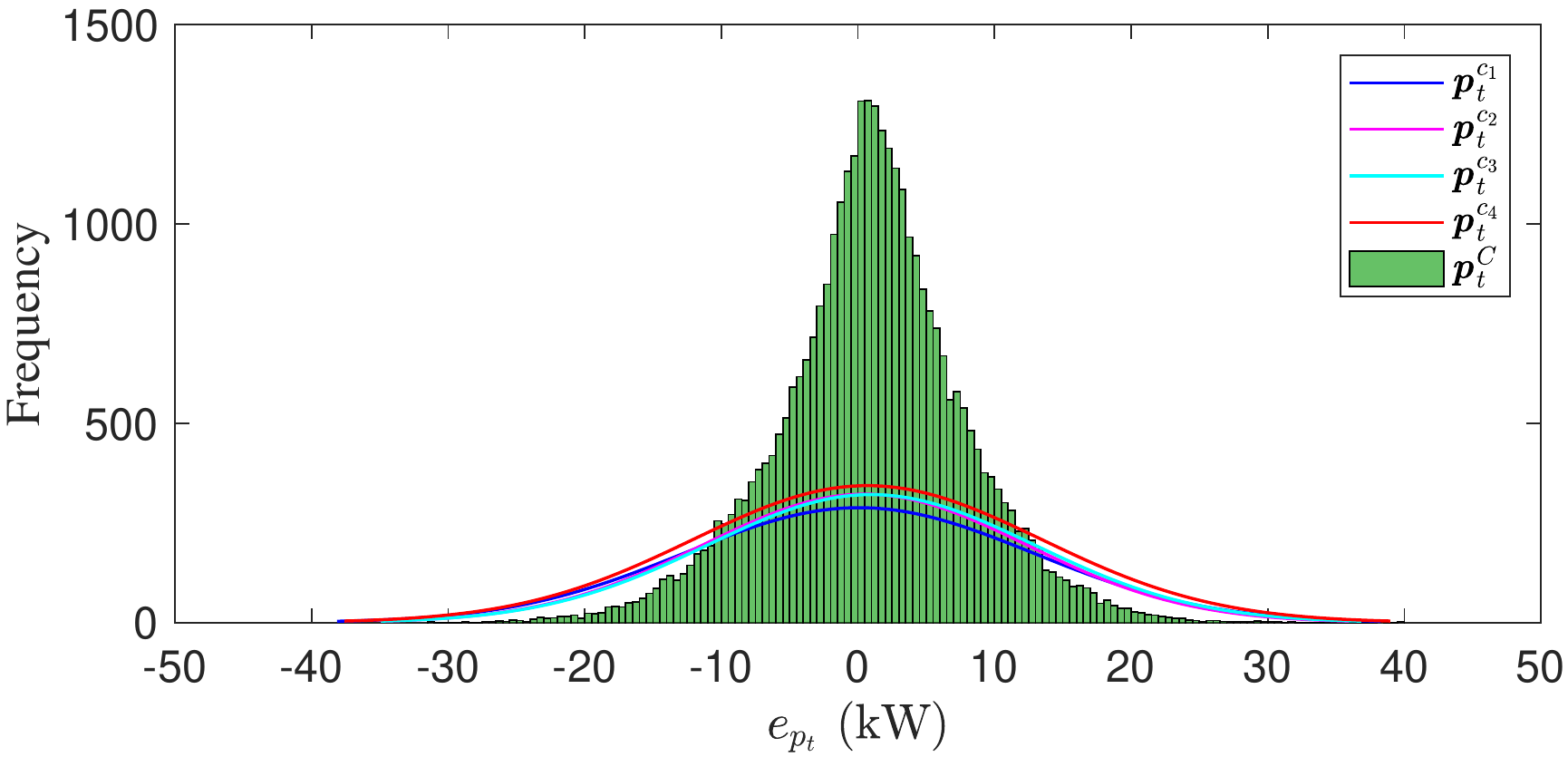}
}
\caption{Error distribution under composite exemplar and candidate exemplars.}
\label{fig:residual_comparison}
\end{figure}

To validate RGVP, Fig. \ref{fig:residual_comparison} employs error histograms to demonstrate the performance of the disaggregation process corresponding to RGVP-based composite exemplars and individual candidate exemplars, where $e_{g_t}=g(t)-\hat{g}_(t)$ and $e_{p_t}=p(t)-\hat{p}(t)$ represent solar and demand disaggregation errors, respectively. As can be seen, the error distribution under composite exemplars for both solar and native demand show considerably lower variance (i.e., higher precision), compared to those of candidate exemplars. This implies that using the game-theoretic learning process to construct composite exemplars leads to improvements in the disaggregation accuracy on average.

\begin{figure}[h]
\centering
\subfloat[PV generation\label{sfig:sensi_PV}]{
\includegraphics[width=0.8\linewidth]{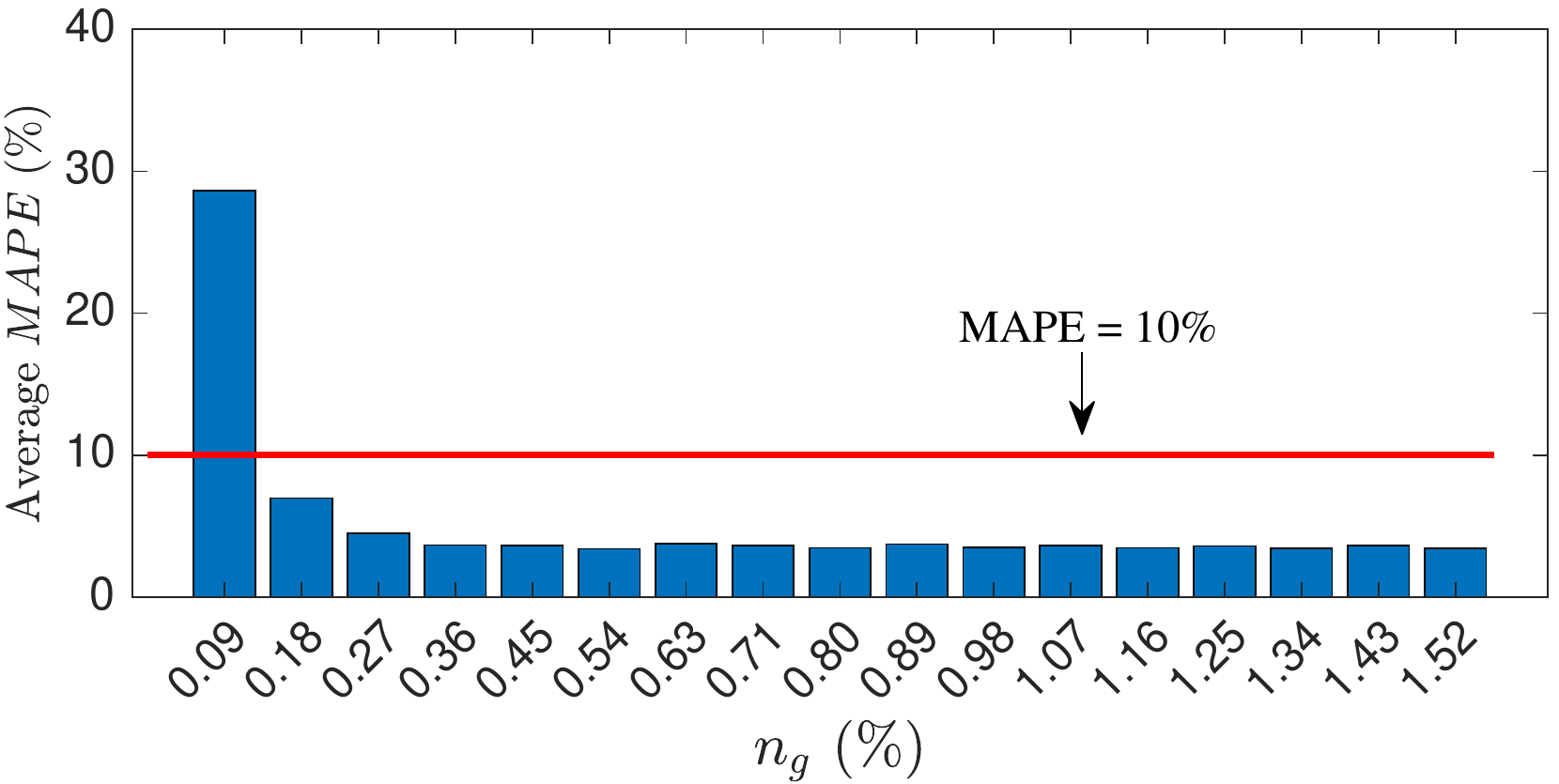}
}
\hfill
\subfloat[Native demand\label{sfig:sensi_demand}]{
\includegraphics[width=0.8\linewidth]{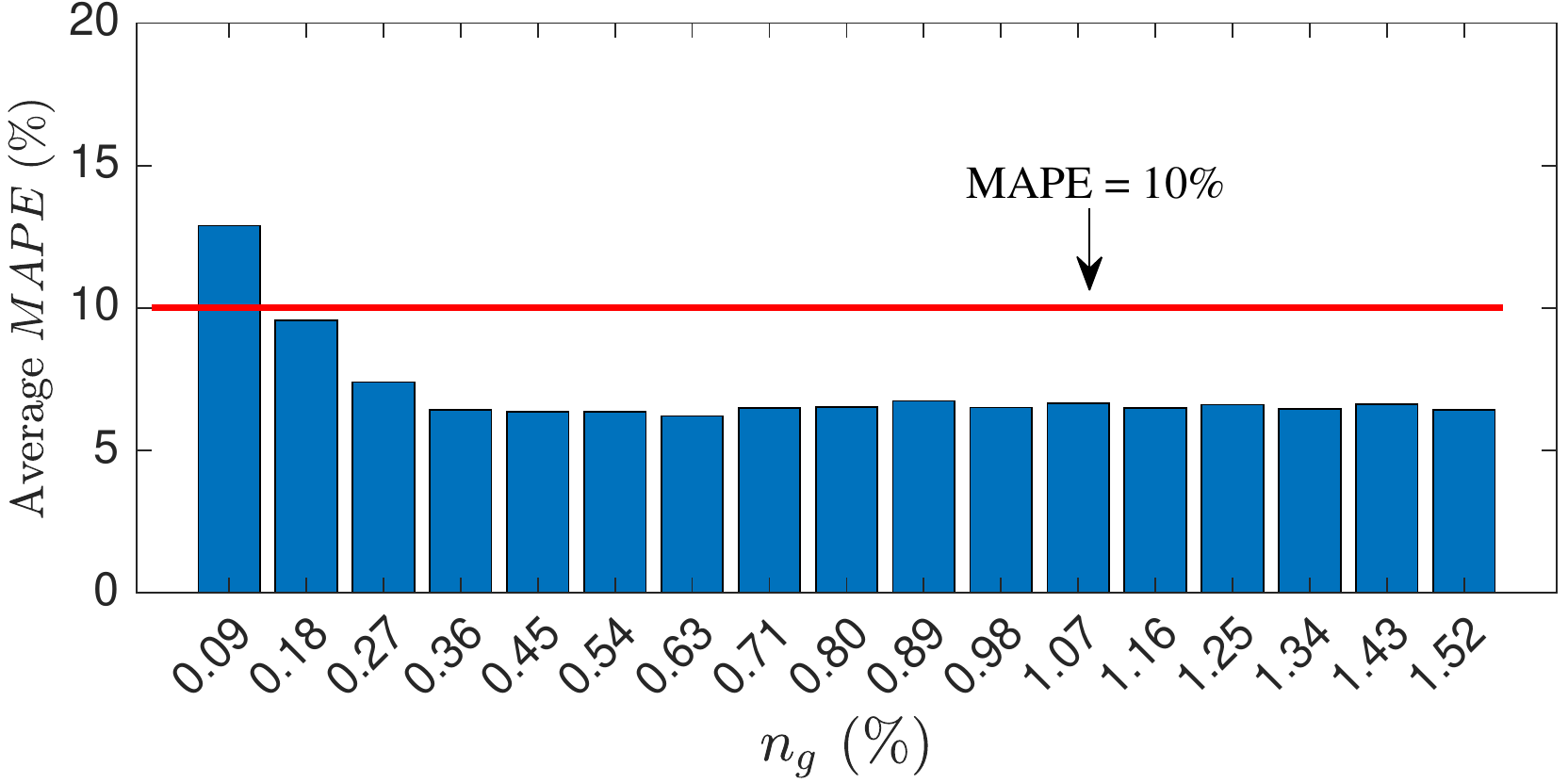}
}
\caption{Sensitivity analysis for the number of observable PVs.}
\label{fig:sensitivity_analysis}
\end{figure}

Since the proposed disaggregator depends on exemplary profiles, it is critical to conduct additional numerical analysis to capture the sensitivity of disaggregation accuracy with respect to the number of observable customers in $S_G$ and $S_P$. This has been done by reducing the percentage of observable PVs, $n_g$, from 1.5\% to 0.1\% in 0.1\% steps (1 PV per step) and performing disaggregation at each step. The average $MAPE$s of disaggregated PV generation and native demand are plotted against the percentage of observable PVs, as shown in Fig. \ref{fig:sensitivity_analysis}. As can be seen in Fig. \ref{sfig:sensi_PV}, once $n_g$ drops below 0.3\%, the average $MAPE$ significantly increases. In the worst case, when only 1 customer is observable ($n_g\approx$ 0.1\%), the average MAPE is about 28\%. This is consistent with expectations: (1) Only a single observable PV cannot represent all other PVs, since the PV panel orientation has significant impact on PV generation profile; (2) As $n_g$ increase from 0.1\% to 0.3\%, the average $MAPE$ significantly decreases, because the unobservable PVs can be better represented using more diverse PV generation profiles; (3) When $n_g$ is larger than 0.3\%, further increase in $n_g$ does not lead to any noticeable accuracy improvements since the redundant exemplary PV generation profiles do not contain much additional information. In Fig. \ref{sfig:sensi_demand}, a similar decreasing trend in average $MAPE$ for the disaggregated native demand can also be observed, which is consistent with the constraint in Equation (\ref{eq:overall_b}). Also, we have performed numerical sensitivity analysis to capture the impact of number of observable customers in $S_P$ on disaggregation accuracy. Similar to the case of $S_G$, as the number of customers with observable native demand increases the disaggregation accuracy improves. In our case studies, the minimum required customers with observable demand is 18\%. Note that these numbers are case-dependent, and vary for different customer behaviors in different regions.

\begin{figure}[h]
\centering
\subfloat[PV generation estimation MAPE\label{sfig:PV_generation}]{
\includegraphics[width=0.8\linewidth]{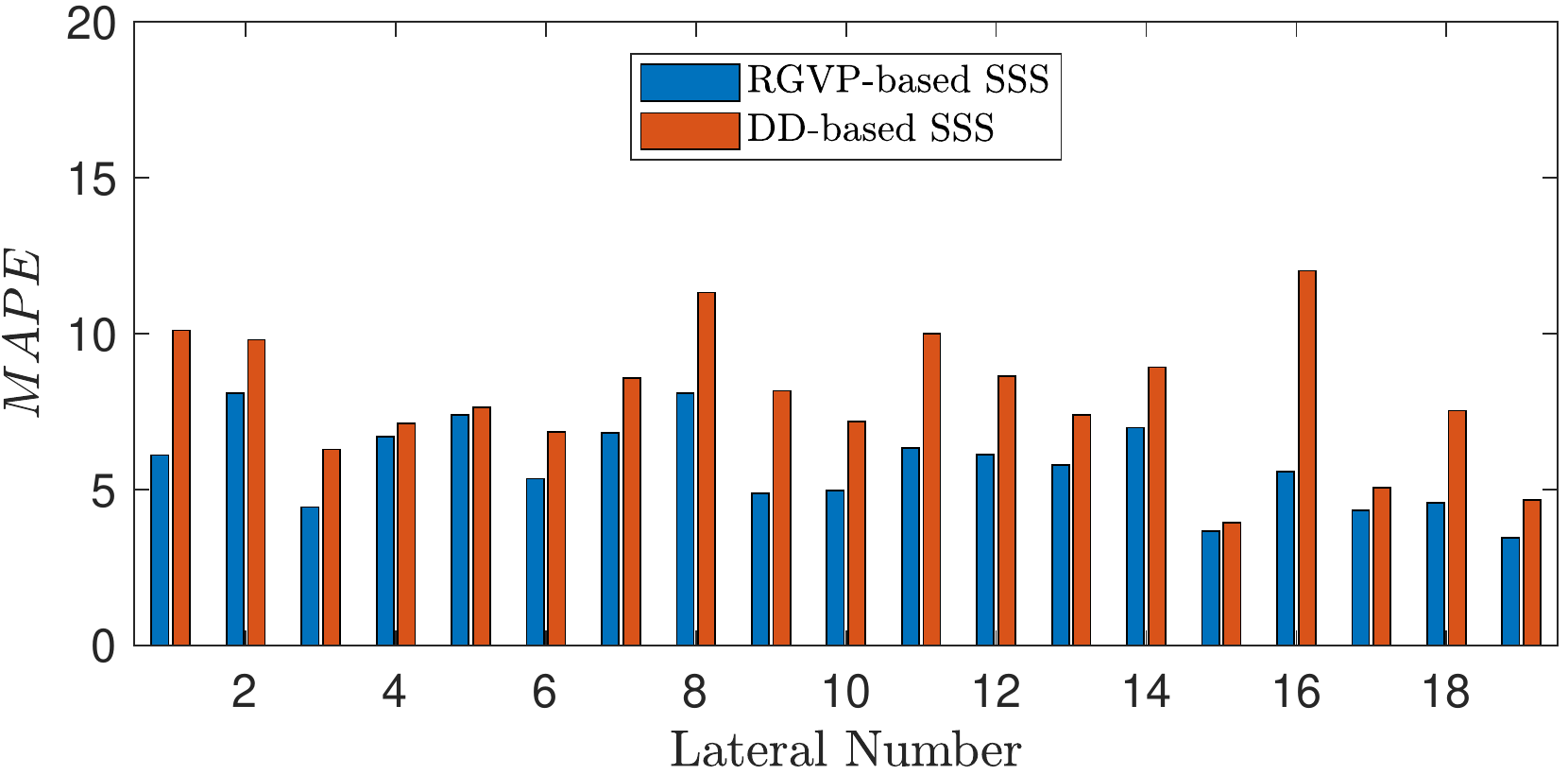}
}
\hfill
\subfloat[Native demand estimation MAPE\label{sfig:native_demand}]{
\includegraphics[width=0.8\linewidth]{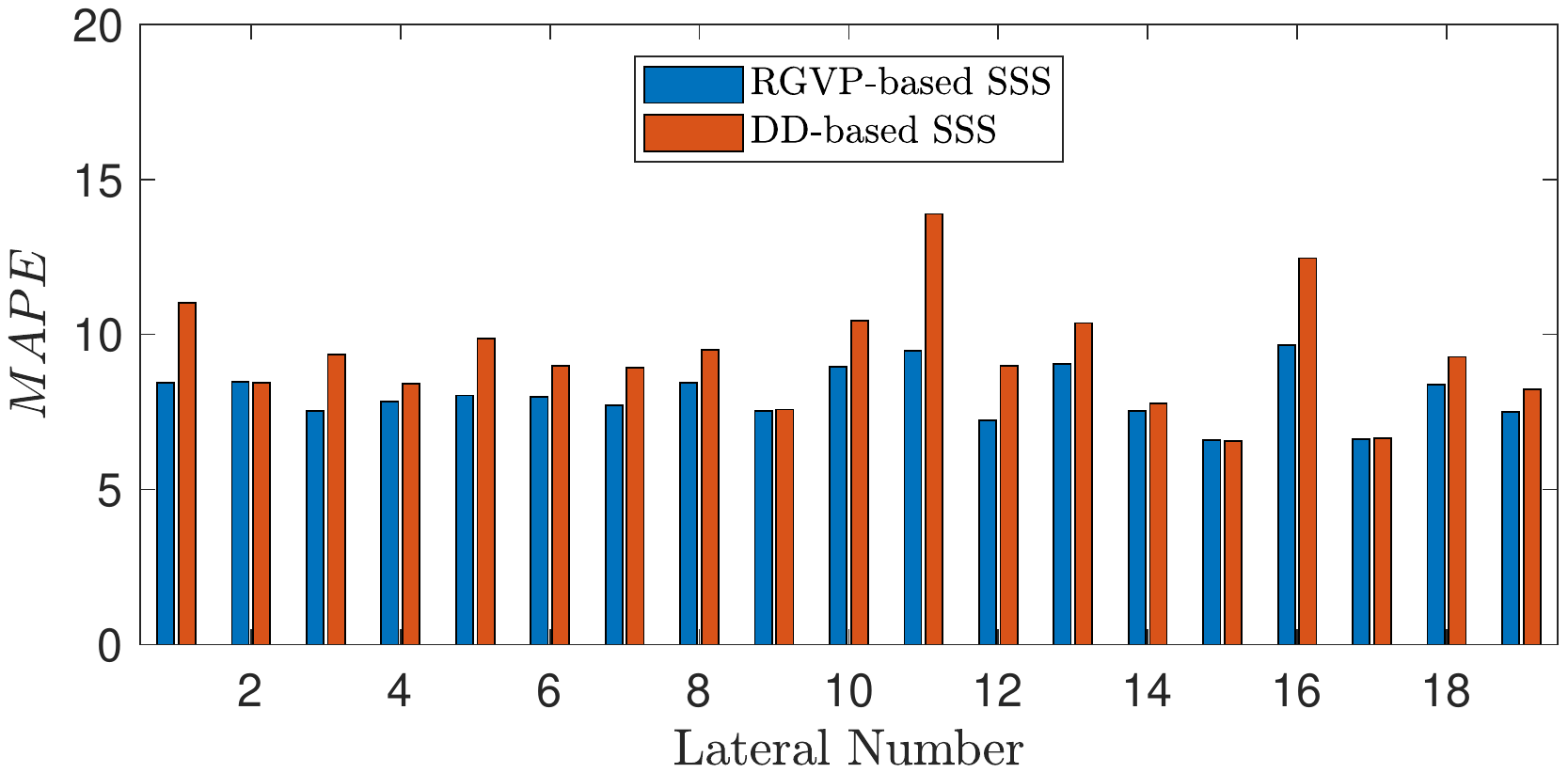}
}
\caption{MAPE comparison using RGVP-based SSS and DD-based SSS.}
\label{fig:MASE_comparison}
\end{figure}

\begin{figure*}[h]
\centering
\subfloat[20\% PVs are in failure condition\label{sfig:adaptability_a}]{
\includegraphics[width=0.95\linewidth]{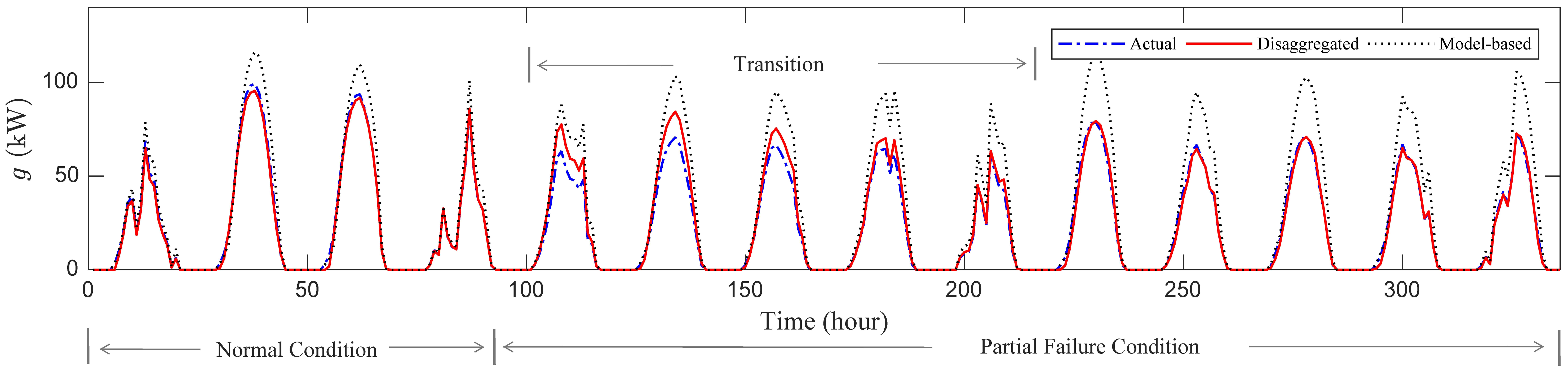}
}
\hfill
\subfloat[40\% PVs are in failure condition\label{sfig:adaptability_b}]{
\includegraphics[width=0.95\linewidth]{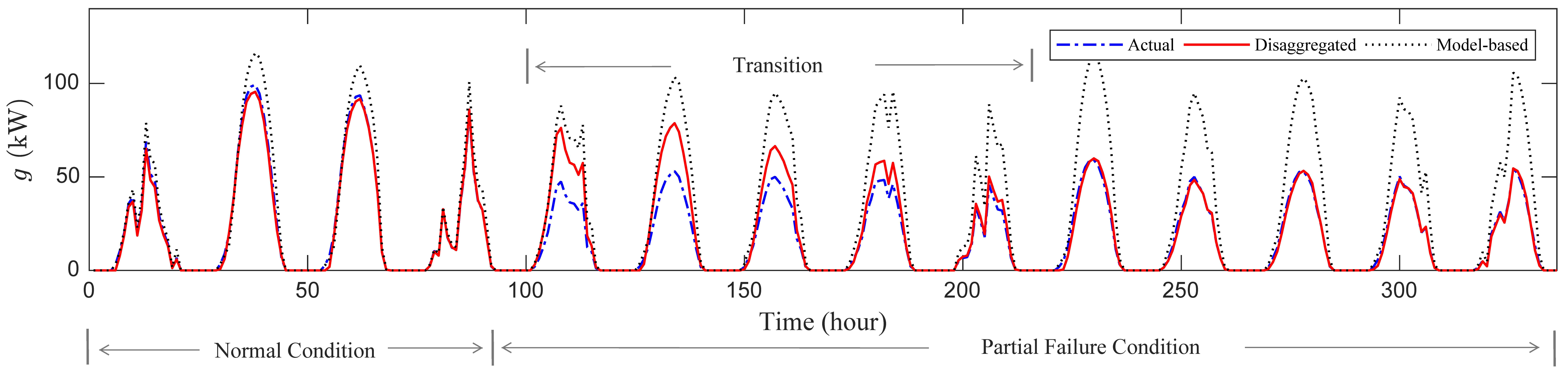}
}
\caption{Adaptability of the proposed data-driven approach.}
\label{fig:adaptability}
\end{figure*}

To further demonstrate the advantage of our proposed approach, we have conducted simulations in two scenarios where a certain percentage of BTM PVs (20\% and 40\%) have stopped running without the utility's knowledge. A model-based method \cite{PV_model_sandia} has been used as a benchmark for comparison with our data-driven technique. In Fig. \ref{fig:adaptability}, the real PV generation curves, the disaggregated curves from the data-driven method and the estimated curves from PV model are plotted with different PV failure percentages for comparison. As can be seen, regardless of the percentage of faulty PVs, the model-based method cannot detect PV failure and cannot adjust to PV generation estimation. This inflexibility is due to the model's inability to adapt to changes in system conditions, which are BTM and unknown. In contrast, our data-driven approach displays high adaptability to PVs' failure conditions. Specifically, although the real PV generation decreases to a lower level due to PV failure, our data-driven disaggregator can track this change after a short transition phase. The small difference between the real and the estimated curves, both before the PVs' failure and after the transition, demonstrates satisfying disaggregation accuracy. The adaptability of the proposed disaggregator can also be extended to unauthorized PV installation and expansion, since essentially both partial failure and installation can be translated into changes in BTM capacity of generation.

An alternative approach to the proposed RGVP is to directly perform PV generation disaggregation using candidate exemplars without developing composite exemplars. For abbreviation, we denote this approach as ``direct disaggregation-based SSS" (DD-based SSS). The solar-demand disaggregation accuracy of the DD-based SSS and the proposed RGVP-based SSS are shown in Fig. \ref{fig:MASE_comparison}. It can be seen that the proposed RGVP-based SSS outperforms DD-based SSS in terms of $MAPE$. The reason for this better performance is that the RGVP-based method can identify the candidate exemplars that are highly correlated with the BTM real load/solar powers.

\begin{figure}
\centering
\includegraphics[width=0.85\linewidth]{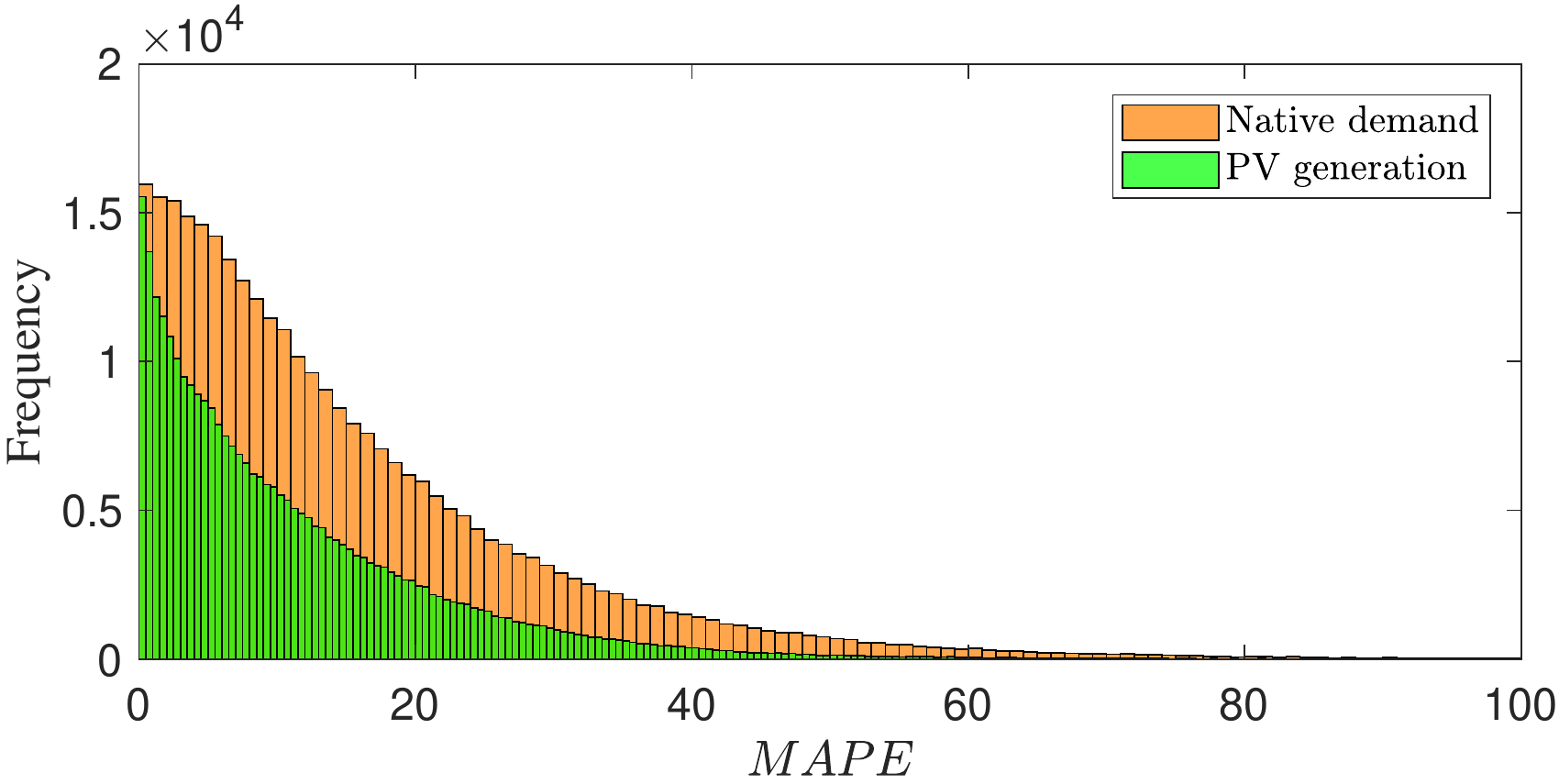}
\caption{Estimation MAPE distribution for PV generation and native demand at secondary transformer level.}
\label{fig:MASE_trans}
\end{figure}

The proposed BTM PV generation and native demand disaggregation approach is also applied to secondary transformers. Fig. \ref{fig:MASE_trans} shows the transformer-level disaggregation $MAPE$ distribution. It can be seen that the proposed method is able to achieve an average solar disaggregation $MAPE$ of 10.0\%, with an average native demand disaggregation $MAPE$ of 15.0\%. As can be seen, disaggregation at transformer-level results in higher residuals compared to lateral-level due to increased grid-edge demand volatility.

\section{Conclusion}\label{sec:conclusion}
This paper presents a non-intrusive novel RGVP-based approach to disaggregate BTM solar generation from the net demand. The proposed method employs the data of fully observable customers to identify typical demand/generation patterns, and optimally combines these patterns to improve disaggregation performance over time. We have used real smart meter data and practical distribution system models from our utility partners to show that this technique is able to enhance solar disaggregation accuracy by adaptively updating the estimator's response to volatile BTM resources. This can enhance utilities' situational awareness of grid-edge resources without incurring extra metering investment costs. The key findings of the paper are summarized as follows:
\begin{itemize}
\item Using real smart meter data, we have observed that: the native demand of any two sizable groups of customers are highly correlated; any two PV generation profiles with similar orientations are significantly correlated; the correlation between PV generation and native demand is insignificant. Based on these three observations, we have proposed a novel data-driven PV generation disaggregator which only relies on utilities' existing smart meter data to separate native demand and BTM solar power.
\item  Numerical experiments have demonstrated that our approach can accurately perform solar generation disaggregation without knowing the specific parameters of BTM PV array and inverters, or weather information. This gives our method a considerable edge over parameter-dependent model-based techniques.
\item  The numerical experiments have also verified that the proposed disaggregator shows strong robustness and adaptability to unobservable BTM abnormal events, such as PV failure and unauthorized PV array installation/expansion.
\end{itemize}

The proposed approach shows satisfactory performance on feeder/lateral-level PV generation disaggregation in which our utility partners have shown great interest; however, when applied to individual customers' data, the disaggregation accuracy declines. In the future, we intend to address this challenge.


\ifCLASSOPTIONcaptionsoff
  \newpage
\fi



\bibliographystyle{IEEEtran}
\bibliography{IEEEabrv,./bibliographies/reference}   

\end{document}